\documentclass[twocolumn, showpacs, amsmath, amssymb, prb]{revtex4}

\usepackage{amssymb}
\usepackage{amsmath}
\usepackage{amsfonts}
\usepackage{graphicx}

\makeatother

\newcommand{\be}{\begin{equation}}
\newcommand{\ee}{\end{equation}}
\newcommand{\ba}{\begin{eqnarray}}
\newcommand{\ea}{\end{eqnarray}}

\begin{document}

\title{Theory for absorption of ultrashort laser
       pulses by spheroidal metallic nanoparticles}

\author{Nicolas I.~Grigorchuk \footnote{email: ngrigor@bitp.kiev.ua}}

\affiliation{Bogolyubov Institute for
Theoretical Physics, National Academy of Sciences of Ukraine, \\
 14-b Metrologichna Str., Kyiv-143, Ukraine, 03680}

\author{ Petro M. Tomchuk \footnote{email: ptomchuk@iop.kiev.ua}}

\affiliation{Institute for Physics, National Academy of Sciences of Ukraine, \\
 46, Nauky Ave., Kyiv-28, Ukraine, 03680}

\begin{abstract}
The theory for the electric and magnetic fields energy absorption
by small metallic particles subjected to the irradiation
by ultrashort laser pulses of different duration in the region of surface
plasmon excitation is developed. For the particles of the oblate or prolate
spheroidal shape there has been found the dependence of the absorbed energy
on a number of factors, including a particle radius, a degree of the shape
deviation from a spherical one, a pulse duration, the orientation of the
magnetic field upon particle, the magnitude of carrier frequency, and the
value of a shift of the carrier frequency of a laser ray from the frequency
of the surface plasmon excitation in a spherical particle.
An appreciable absorption grows at the length of free electron
pass large compared to the particle size is established.
The phenomenological and kinetic approach is compared each with other.
\end{abstract}


\date{\today}

\maketitle


\newpage
\section{INTRODUCTION}

Metallic nanoparticles (MNs) are interesting objects from the
viewpoints of physics of condensed state and practical applications.
An increase in the local electric fields in the close vicinity of
nanoparticles allows the utilization of the MNs as biological markers
\cite{BTM} and can to lead to entirely new types of biological
sensors.\cite{Ali} The modern optical investigations make it possible
to test the optical response of a separate nanoparticle and, thus, to
study the properties of separate nanoobjects.\cite{SL} This opens the
new opportunities for the direct probing of the many-electron dynamics
in confined systems. The remarkable results were obtained recently
in works.\cite{AA}-\cite{Per} In recent years, the
ultrafast dynamics of electrons in MNs
attracts the permanent attention of experimenters.
Nanostructures in general have been widely used in modern devices of
high-speed electronics and optoelectronics. Owing to the recent
achievements in the ultrafast spectroscopy technique, it has become
possible to monitor the femtosecond dynamics of the electron gas
confined in separate MNs.\cite{BHM} The use of short-duration pulses
makes it feasible to study the dynamics of fast processes occurring
in atoms, molecules, and solids. Pico- and femtosecond resolutions
allow to study of vibrational and rotational intramolecular motions,
the dynamics of carriers in semiconductor nanostructures, the phase
transitions in solids, the processes of formation and breakdown of
chemical bonds, etc. \cite{AVC,Zhe} The possibility to generate powerful
femtosecond pulses has lead to the appearance of a new, rapidly
developing branch -- the desk-mounted high-energy physics.

When a MN is placed in the field of a monochromatic electromagnetic
(EM) wave and the wavelength is far greater than the particle size,
the absorption is observed associated both with the separate and
collective kinds of electron motion. The input to the separate
absorption gives as an electric component of an EM wave (electric
absorption), so the magnetic component (magnetic absorption) as
well. The collective electron motion become apparent in so-called
surface resonances connected with the excitation of plasmon
vibrations (plasmons) of an electron gas.\cite{KV,Hee} There are three
such resonances in an ellipsoidal particle and only one in a spherical
particle. In the case of a monochromatic wave, the collective
contribution becomes dominant provided that the wave frequency is
close to one of the frequencies of plasma resonances.\cite{Hao} Numerous
applications in decorative painting \cite{KV} and optical devices \cite{Hau}
are based just on the resonances of surface plasmons. However, in spite
of the appreciable theoretical efforts, \cite{WMW} a body of problems has
remained unsolved hitherto.

In previous works \cite{TG} we had shown, that by exposure of
metallic nanoparticle to EM wave, the frequency of which is
far from plasmon resonances, a separate mechanism becomes actually
in absorption. Depending on the shape and size of the particle, as well
as on the frequency and polarization of wave, it was shown that, the
magnetic absorption can be higher or less than the electric one.
The situation is drastically change with using of an ultrashort laser
pulses for irradiation, because they contains almost all harmonics.
The shorter the pulse, the greater is set of its harmonics.
Thus, the ultrashort pulses can excite all the plasma resonances.
This is associated with the fact that, in such an ultrashort pulse,
one can always find the harmonics which correspond to the resonance
frequencies of surface plasmons. From the other side, the frequency
distribution is very nonuniform (Gaussian) upon the spectra of an
ultrashort pulses. How these factors can change the relative input of
an electric and magnetic components in absorption, we intend to study
below. Another aspect influencing the absorption is associated with the
shape of a particle. Earlier, we showed that, in the case of nonspherical
MNs, the optical conductivity, which determines both the electric
absorption and the half-width of plasma resonances, becomes the
tensor quantity.\cite{TG} As far as we know, such a feature in
absorption of ultrashort laser pulses by nonspherical MNs was not
investigated in details.

In this paper we study for the first time, the peculiar features of
the absorption by the MNs of ultrashort pulses, deduced by an electric
and magnetic components of EM wave. Such a problems remains less studied,
especially for the particles of a nonspherical shape, whose dimensions
are less or exceed the electron free path in it, and
have attracted a lot of interest of experimenters in recent years.
In particular, we will find: (i) the energy absorbed by MNs from
ultrashort laser pulses of a different duration; (ii) its dependence
on the particle shape, frequency of a carrier wave, pulse duration, and
particle sizes; (iii) the shift of the resonance frequency of the
absorption with respect to the plasmon frequency in a spherical MN.

The rest of the paper is organized as follows. The model and an
initial foundations of problem are presented in Sec. II. Section III
contains the study of a plasmon resonances in the MNs. In section IV
the study of magnetic absorption is demonstrated using both the
phenomenological and the kinetic approaches,
and Sec. V contains the conclusions.

\section{MODEL AND STARTING POSITIONS}

Consider a case where an MN is irradiated by a laser pulse, whose
electric field is given as

\[
 \mathrm{\mathbf{E}}(\mathrm{\mathbf{r}},t)\!=\!\mathrm{\mathbf{E}}_{0}\exp
  \left[-\Gamma ^{2}\left( {\!t-{\frac{{\mathrm{\mathbf{k}}_{\mathrm{\mathbf{0}}}
   \mathrm{\mathbf{r}}}}{{\omega _{\mathrm{\mathbf{0}}}}}}}\right)^{2}\right]
    \!\cos \!{\left[ {\omega _{0}\left( {t-{\frac{{\mathrm{\mathbf{k}}_{\mathrm{
     \mathbf{0}}}\mathrm{\mathbf{r}}}}{{\omega _{\mathrm{\mathbf{0}}}}}}}\right) }
      \right] ,}
       \]\vskip-7mm
        \begin{equation}
         \label{eq1}
          \end{equation}
where $\Gamma$ is the quantity reciprocal to the pulse duration,
$\omega_{0}$ is the carrier frequency of an EM wave,
$\left\vert \mathrm{\mathbf{k}}_{\mathrm{\mathbf{0}}}\right\vert =\omega _{0}/c$,
and $\mathrm{\mathbf{E}}_{0}$ is the maximal value of an electric field in a pulse.
In addition to the electric component, the field of a laser pulse contains also
the magnetic one which is associated with the former through the
corresponding Maxwell equation
\begin{equation}
 \mathrm{rot}\;\mathrm{\mathbf{E}}(\mathrm{\mathbf{r}},t)=-{\frac{1}{c}}\,{
  \frac{{\partial }}{{\partial t}}}\mathrm{\mathbf{H}}(\mathrm{\mathbf{r}},t).
   \label{eq2}
    \end{equation}
We set $\mathrm{\mathbf{E}}(\mathrm{\mathbf{r}},t)$ in the form
(\ref{eq1}) and will find the magnetic component of a pulse, by
using Eq.~(\ref{eq2}). Relation (\ref{eq2}) takes the simplest form
if one operates with the Fourier components of the above quantities.
Actually, for an electric field, e.g., we obtain from Eq. (\ref{eq1})
\ba
 \mathrm{\mathbf{E}}(\mathrm{\mathbf{r}},\omega ) &=&
  \int\limits_{-\infty}^{\infty }\mathrm{\mathbf{E}}(\mathrm{\mathbf{r}},t) e^{i\,\omega \,t} dt=
   \mathrm{\mathbf{E}}_{\mathrm{\mathbf{0}}}\frac{\sqrt{\pi}}{2\;\Gamma}e^{i\,
    \mathrm{\mathbf{k}}_{\mathrm{\mathbf{0}}}\mathrm{\mathbf{r}}\;\left(
     \frac{\omega}{\omega _{0}}\right) }
      \nonumber\\&\times&
       \left[ \exp \left( -{\frac{{(\omega -\omega _{0})^{2}}}{{4\Gamma^{2}}}}
        \right) +\exp \left( -{\frac{{(\omega +\omega_{0})^{2}}}{{4\Gamma ^{2}}}}\right) \right] .
         \label{eq3}\nonumber \\
          \ea
Carrying out the Fourier transformation in Eq.~(\ref{eq2}), one finds connection
between Fourier components of the magnetic and electric fields as
\[
 \mathrm{\mathbf{H}}(\mathrm{\mathbf{r}},\omega)=
  \frac{c}{i\omega}\mathrm{rot}\;\mathrm{\mathbf{E}}(\mathrm{\mathbf{r}},\omega).
   \]
Using the last part of Eq.~(\ref{eq3}), we find for magnetic field

\ba
 {\rm {\mathbf H}}({\rm {\bf r}},\omega ) &=& \frac{\sqrt {\pi}}{2\;
   \Gamma}({\rm {\bf m}}\times \rm {\bf E}_{\rm {\bf 0}})
    \nonumber\\
    &\times& \left( {e^{- \frac{(\omega -\omega_{0} )^2}{4\Gamma^2}}
     + e^{-\frac{(\omega + \omega _{0} )^2}{4\Gamma^2}}}
       \right)\;e^{i\,{\rm {\bf k}}_{{\rm {\bf 0}}} {\rm {\bf r}}\;
        \left(\frac{\omega}{\omega_0} \right)},
         \label{eq4}
          \ea
where
$\mathrm{\mathbf{m}}=\mathrm{\mathbf{k}}_{\mathrm{\mathbf{0}}}/k_{0}$
is the unit vector directed along the EM wave propagation.

The Fourier component of the electric field amplitude takes the
simplest form in the limit $\Gamma \rightarrow 0$:
\begin{equation}
 \mathrm{\mathbf{E}}(\mathrm{\mathbf{r}},\omega )_{\Gamma
  \rightarrow 0}=\pi\mathrm{\mathbf{E}}_{\mathrm{\mathbf{0}}}e^{i\mathrm{
   \mathbf{k}}_{\mathrm{\mathbf{0}}}\mathrm{\mathbf{r}}(
    \frac{\omega}{\omega _{0}})}{\left[ {\delta(\omega -
     \omega _{0})+\delta (\omega +\omega _{0})}\right] }.
      \label{eq5}
       \end{equation}

Whereas the electric field of a laser pulse creates the potential
electric field
$\mathrm{\mathbf{E}}_{\mathrm{in}}(\mathrm{\mathbf{r}},t)$ inside an
MN, the magnetic field induces the eddy current field
$\mathrm{\mathbf{E}}_{\mathrm{ed} }(\mathrm{\mathbf{r}},t)$. To
calculate the energy absorbed by an MN, we should know the inner
fields $\mathrm{\mathbf{E}}_{\mathrm{in}}(\mathrm{\mathbf{r}},t) $
and $\mathrm{\mathbf{E}}_{\mathrm{ed}}(\mathrm{\mathbf{r}},t)$. To
find these quantities, we pay attention to the following feature of
the coordinate dependence of the Fourier components (\ref{eq3}) and
(\ref{eq4}): when the characteristic dimension $R$ of a nanoparticle
is such that
\begin{equation}
 k_{0}R\ll 1,
  \label{eq6}
   \end{equation}
i.e. the wavelength of the carrier wave is far greater than the MN
dimension, the coordinate dependence of the Fourier components $\mathrm{%
\mathbf{E}}(\mathrm{\mathbf{r}},\omega )$ and $\mathrm{\mathbf{H}}(\mathrm{%
\mathbf{r}},\omega )$ can be neglected inside the particle. This
means that,
in order to determine the inner fields inside an MN, provided that inequality
(\ref{eq6}) is satisfied, we can set the Fourier components of these
fields such as they would be in the spatially uniform fields:
 \begin{equation}
  \mathrm{\mathbf{E}}(\mathrm{\mathbf{r}},\omega )\rightarrow
   \mathrm{\mathbf{E}}(0,\omega )~~{\mathrm{ and}}~~\mathrm{\mathbf{H}}(
    \mathrm{\mathbf{r}},\omega )\rightarrow
     \mathrm{\mathbf{H}}(0,\omega ).
      \label{eq7}
       \end{equation}
In the case of an nonspherical MN which has, for example, the form of
an ellipsoid, this allows us to write the components of the inner
electric field as
\begin{equation}
 [E_{{in}}(\omega )]_{j}\approx {\frac{{E_{j}^{(0)}(0,
  \omega )}}{{1+L_{j}{\left[ {\varepsilon (\omega )-1}\right] }}}},
   \label{eq8}
    \end{equation}
and the magnetic eddy current field as
\ba
 [E_{\mathrm{ed}} ({\rm {\bf r}},\omega )]_{x} & \approx & i{\frac{{\omega}}{c}}R_{x}^{2}
  \nonumber\\
 & \times &
    \left[ {{\frac{{H_{y}^{(0)}(0,\omega ) }}{{R_{z}^{2} + R_{x}^{2}}}} z -
     \frac{{H_{z}^{(0)} (0,\omega )}}{{R_{x}^{2} + R_{y}^{2}}} y  } \right]
      \label{eq9}
       \ea
\noindent by analogy with the procedure described in Ref. [\onlinecite{GT}].
The other components of the eddy current field can be obtained from Eq.~(\ref{eq9})
by means of a cyclic permutation of indices. The notations in formulas
(\ref{eq8}) and (\ref{eq9}) are as follows: $L_{j}$ ($j=x,y,z$) are the
geometrical factors, also known as the depolarization factors,\cite{LL}
$\varepsilon (\omega )=\varepsilon ^{\prime }(\omega )+i\varepsilon
^{\prime \prime }(\omega )$
is the frequency-dependent complex permittivity of the particle, $%
R_{x}$, $R_{y}$, $R_{z}$ are the ellipsoid semiaxes in the $x$, $y$,
and $z$ directions, respectively.

The linear dependence of the eddy current field on coordinates can
be easy seize from Eq.~(\ref{eq2}), which determined it. If one
perform the time Fourier-transform of Eq.~(\ref{eq2}), than the
right-hand side of this equation [as it is seen from both the Eq. (\ref{eq4})
and condition (\ref{eq6})], one can consider as permanent,
coordinates-independent quantity. This means that
$rot\;{\rm {\bf E}}_{e d}({\rm {\bf r}},\omega ) = const$.
The last equality is satisfy only if
\begin{equation}
[{\rm {\bf E}}_{ed} ({\rm {\bf r}},\omega )]_{j} ={
   \sum\limits_{k = 1}^{3} {\alpha _{jk} (\omega )\,x_{k}}}  ,
    \label{eq10}
     \end{equation}

\noindent e.g., the curly field depends on coordinates linearly.
Here $\alpha $ is some matrix does not depended on the ${\rm {\bf
r}}$; its components will be further specified.

It is seen from expressions (\ref{eq8}) and (\ref{eq9}) that the
uniform external electric field induces also the uniform electric
potential field inside the ellipsoidal MN, whereas the uniform
external magnetic field generates the
coordinate-dependent eddy current field. The inner fields $\mathrm{\mathbf{E}%
}_{\mathrm{in}}(\mathrm{\mathbf{r}},t)$ and $\mathrm{\mathbf{E}}_{{ed}}(%
\mathrm{\mathbf{r}},t)$ induce the currents inside the MN with the
corresponding
densities $\mathrm{\mathbf{j}}_{{in}}(\mathrm{\mathbf{r}},t)$ and $%
\mathrm{\mathbf{j}}_{\text{ed}}(\mathrm{\mathbf{r}},t)$.\ As a
result, the particle absorbs the energy of the EM field of an
incident laser wave. The total absorbed energy is sum of electric
and magnetic absorption and can be presented as\cite{LL}
\ba
 w_t &=& {\int\limits_{ - \infty} ^{\infty}  {dt\;W(t)}}  = w_{e}+ w_{m} =
  {\int\limits_{ - \infty}^{\infty} {dt
   \;{\int\limits_{V} {d{\rm {\bf r}}}}} }
    \nonumber\\
   & \times &
      \left[{\rm Re}\,{\rm {\bf j}}_{e}({\rm {\bf r}},t)
       \;{\rm Re}\,{\rm {\bf E}}_{\mathrm{in}}^{\ast} ({\rm {\bf r}},t)\right. \nonumber\\
   & + & \left.
       {\rm Re}\,{\rm{\bf j}}_{\mathrm{ed}} ({\rm {\bf r}},t)\;
         {\rm Re}\,{\rm {\bf E}}_{\mathrm{ed}}^{\ast} ({\rm {\bf r}},t)\right],
          \label{eq11}
           \ea
where the integration should be carried out over the whole particle
volume. Here, $W(t)$ is the absorbed power. Depending on whether the
absorption is brought about by
$\mathrm{\mathbf{E}}_{\mathrm{in}}(\mathrm{\mathbf{r}},t)$ or
$\mathrm{\mathbf{E}}_{\mathrm{ed}}(\mathrm{\mathbf{r}},t)$, it is
called the electric or magnetic absorption, respectively. In view of the equality
${\rm {\bf j^*}}(\omega)={\rm {\bf j}}(-\omega)$ and

\[
{\int\limits_{-\infty }^{\infty }{\mathrm{\mathbf{j}}_{e}(\mathrm{\mathbf{r}}
,t)\;\mathrm{\mathbf{E}}_{\mathrm{in}}^{\ast }(\mathrm{\mathbf{r}},t)\;dt=}}{
\int\limits_{-\infty }^{\infty }{\mathrm{\mathbf{j}}_{e}(\mathrm{\mathbf{r}}
,\omega )\;\mathrm{\mathbf{E}}_{\mathrm{in}}^{\ast }(\mathrm{\mathbf{r}}
,\omega )\;{\frac{{d\omega }}{2\pi}}}},
\]
which follows, in particular, from the Parseval relation for the
Fourier integral,\cite{Bre} expression (\ref{eq11})\ can be rewritten as
\ba
 &w_t&= {\frac{{1}}{{4\pi}} }{\int\limits_{ - \infty} ^{\infty}{d\omega
  \;{\int\limits_{V} {d{\rm {\bf r}}}}} }
   \nonumber\\
  & \times &
     \left[ {{\rm {\bf j}}_{e} ({\rm {\bf r}},\omega )
      \;{\rm {\bf E}}_{\mathrm{in}}^{\ast} ({\rm {\bf r}},\omega ) +
      {\rm {\bf j}}_{\mathrm{ed}} ({\rm {\bf r}},\omega)
        \;{\rm {\bf E}}_{\mathrm{ed}}^{\ast}  ({\rm {\bf r}},\omega )}+c.c.
         \right].
          \label{eq12}
           \ea

The values of $\mathrm{\mathbf{E}}_{\mathrm{in}}(\omega )$ and $\mathrm{%
\mathbf{E}}_{\mathrm{ed}}({\mathrm{\mathbf{r,}}}\omega )$ can be
calculated from formulas (\ref{eq8}) and (\ref{eq9}), respectively.
Thus, the problem remaining to be done is to find the Fourier
components of the current densities
${\mathrm{\mathbf{j}}_{\mathrm{in}}(\mathrm{\mathbf{r}},\omega )}$
and ${\mathrm{\mathbf{j}}_{\mathrm{ed}}(\mathrm{\mathbf{r}},\omega
)}$. In the
general case, the current produced by the inner fields $\mathrm{\mathbf{E}}_{%
\mathrm{in}}(\omega )$ and $\mathrm{\mathbf{E}}_{\mathrm{ed}}({\mathrm{\mathbf{r,%
}}}\omega )$ at a point $\mathrm{\mathbf{r}}$ of the particle can be
expressed as the integral over all values of electron velocities ${\bf v}$
\begin{equation}
 \mathrm{\mathbf{j}}(\mathrm{\mathbf{r}},\omega )=2e\left({{\frac{{m}}{{2\pi
  \hbar }}}}\right) ^{3}{\int\limits_{-\infty}^{\infty }{d^{3}{ v}
   \;{\rm {\bf v}}\;f_{1}(\mathrm{\mathbf{r}},{\rm {\bf v}} ,\omega )}}.
    \label{eq13}
     \end{equation}
Here, $v=|{\bf v}|$, $f_{1}(\mathrm{\mathbf{r}},{\rm {\bf v}},\omega )$ is the
Fourier component of the nonequilibrium distribution function
usually considered as an addition to the equilibrium Fermi distribution function
$f_{0}(\varepsilon )$ which depends only on the electron kinetic energy
$\varepsilon$. Usually, the way to calculate the function
$f_{1}(\mathrm{\mathbf{r}},{\rm {\bf v}} ,\omega )$ consists in the solution
of the corresponding linearized Boltzmann kinetic equation. As a rule, this
equation is written for the time-dependent distribution function
(see, for example, Ref. \onlinecite{GT}). Performing the Fourier
transformation of this equation with the use of expression (\ref{eq3}) and
\[
 f_{1}(\mathrm{\mathbf{r}},\mathrm{\mathbf{v}},\omega)={\int
  \limits_{-\infty }^{\infty}{f_{1}(\mathrm{\mathbf{r}},
   \mathrm{\mathbf{v}},t)\;e^{i\,\omega \,t}\,}}dt,
    \]
we obtain equation for $f_{1}$:%
\ba
 &&(\nu - i\omega )\,f_{1} ({\rm {\bf r}},{\rm {\bf v}},
  \omega ) + {\rm {\bf v}}{\frac{{\partial}} {{\partial {\rm {
   \bf r}}}}}f_{1} ({\rm {\bf r}},{\rm {\bf v}},\omega )
    \nonumber\\
     & + &
      e\;{\rm{\bf v}}\,{\left[ {{\rm {\bf E}}_{\mathrm{in}} (\omega ) + {
       \rm {\bf E}}_{\mathrm{ed}} ({\rm {\bf r}},\omega )}
        \right]}\;{\frac{{\partial }}{{\partial
         \varepsilon}} }f_{0}(\varepsilon ) = 0,
          \label{eq14}
           \ea
where $\nu $ is the collision frequency in the particle bulk. To
move further, we should add the corresponding boundary conditions to
Eq.~(\ref{eq14}). We take assumption of the diffuse reflection of
electrons from the inner surface of a particle
\begin{equation}
 f_{1}(\mathrm{\mathbf{r}},\mathrm{\mathbf{v}},\omega )|_{S}=0,\quad
  \quad \upsilon _{n}<0
   \label{eq15}
    \end{equation}
as the boundary conditions. Here, $\upsilon _{n}$ is the velocity
component normal to the surface $S$. The substantiation of such a
boundary conditions and the solution of (\ref{eq14}) are presented,
in particular, in Ref. \onlinecite{TG}.

Below we consider separately the peculiarities of the electric
absorption when the plasmon resonances are excited by ultrashort
laser pulses and the magnetic absorption take place due to magnetic
component of the laser wave in more details.

\section{PLASMON RESONANCES}

To this point, we have considered the general approach which
includes both the electric and magnetic absorptions. It was stressed
above that the absorption can be either separate or collective.
When an MN is in the field of a EM wave and the wave frequency
is far from the plasmon resonance, the separate mechanism
of absorption prevails. In this case, as was shown in Ref. \onlinecite{TG},
the electric or magnetic absorption can be dominant depending
on a number of factors, namely, on the wave frequency, its
polarization, the MN size and shape. In the present section, we concentrate,
however, on the features of the absorption of ultrashort laser
pulses. In this case, the main contribution to the absorption
results from the plasma resonances, i.e. from the collective
mechanism, which is a constituent part of an electric absorption.
For this reason, we will take into consideration in present section
only that part of absorption in Eq.~(\ref{eq11}), caused by the electric
component of the EM field. To find the total energy of electric
absorption, we will start from expressions (\ref{eq12})
and (\ref{eq13}), and determine the form of the function
$f_{1}(\mathrm{\mathbf{r}},\mathrm{\mathbf{v}},t)$, by solving
Eq.~(\ref{eq14}) with the corresponding boundary conditions. Then,
we will substitute the obtained function into  Eq.~(\ref{eq13}) and
calculate the Fourier components of the current density. Finally, by
taking into account that the Fourier components of the inner field
are described by expression (\ref{eq8}), we will substitute the obtained
Fourier components of the current density into Eq.~(\ref{eq12}) with ${\bf E}_{ed}=0$.
The similar procedure was described in detail in Ref. \onlinecite{TG} and,
thus, we won't repeat it. We only note that the kinetic equation (\ref{eq14})
almost completely coincides with Eq.~(27) of work\cite{TG} with
the only difference: $\mathrm{\mathbf{E}}_{\mathrm{in}}(\omega )$ should
be changed by $\mathrm{\mathbf{E}}_{\mathrm{loc}}(\omega )$.
What is more, the integrand of the integral over frequencies in
Eq.~(\ref{eq12}) formally coincides with the initial expression in
Ref. \onlinecite{TG}. Thus, we can use the results of
Ref. \onlinecite{TG}, in particular formula (75), and write the
total energy of an electric absorption in the form
\begin{equation}
 w_{e}={\frac{{V}}{{2\pi }}}{\sum\limits_{j=1}^{3}{{\int\limits_{-\infty }^{\infty }{
  \omega ^{4}}}}}{\frac{{\sigma _{jj}(\omega )
   \;|E_{j}^{(0)}(0,\omega )|^{2}}}{{\left( {\omega ^{2}-\omega _{j}^{2}}
    \right) ^{2}+(2\gamma_{j}(\omega ))^{2}\omega ^{2}}}}\,d\omega ,
     \label{eq16}
      \end{equation}
where the frequency dependence of $E_{j}^{(0)}(0,\omega )$ is given
by expression (\ref{eq3}). Here,
\begin{equation}
 \gamma _{j}(\omega )=2\pi \; L_{j} \; \sigma _{jj}(\omega )
  \label{eq17}
   \end{equation}
is the half-width of a plasmon resonance, which strongly depends on
the geometrical shape of a particle and is defined in terms of
diagonal components of the optical conductivity tensor $\sigma$.
For the different directions $j$, it is governed by the aforementioned
factor $L_{j}$. For a spheroidal particle $(R_{x}=R_{y}\equiv
R_{\bot })$, these components, for example, are reciprocal to the
square of the frequency at high frequencies and have the form
\begin{equation}
 \sigma _{\binom{\parallel }{\perp }}(\omega )={\frac{{9}}{{32\;
  \pi }}}{\frac{{\upsilon _{\mathrm{F}}}}{{R_{\bot }}}}
   \left( {{\frac{{\omega _{p}}}{{\omega }}}}\right) ^{2}
    \binom{\eta (e_{s})}{\rho (e_{s})},
     \label{eq18}
      \end{equation}
where ${\upsilon _{\mathrm{F}}}$ is the electron velocity on the
Fermi sphere, $\omega _{p}$ is the plasma frequency of electron
oscillations in a metal, and $\eta (e_{s})$ and $\rho (e_{s})$ are
some smooth functions depending on the spheroid eccentricity $e_{s}$.
The analytical dependence of these functions on the ratio between
the spheroid semiaxes as welll as, $e_{s}$ can be found in
Ref. \onlinecite{TG}. For a spherical particle, $\eta =\rho =2/3$.

The half-widths of plasmon resonances are an important characteristic,
since it contains the information about the character of interactions in
the system. Through the interaction between surface plasmons and single-particle
excitations, the electron scattering  gives rise to the decay of
surface plasmons, which reveals itself in the broadening of lines of
the differential transmission spectra. The surface plasmon decay can
occur in different ways, depending on the particle size. Often, the
contribution of the Landau decay \cite{KK} to the linewidth, which is
dominant in the case of a small particle radius $a$ of the order of
5--20 \AA,~\cite{MWJ} is calculated. At greater particle sizes, the
Landau decay competes with the decays caused by other interactions.
Recently, we studied \cite{GT4} how the relation between the transverse
$\gamma_{\perp}$ and longitudinal $\gamma_{\parallel}$ components
of the plasmon resonance half-width depends on the degree of ellipsoid's
oblateness or prolateness for the frequencies that are higher or lower
than the characteristic frequency of electron reflection from the particle walls.

In the case where the plasmon decay is not significant, we can replace
the Lorentzian in expression (\ref{eq16}) by a $\delta $-function as \cite{Arf}
\[
 \lim_{\alpha \rightarrow 0}{\frac{{\alpha }}{{%
  \alpha ^{2}+x^{2}}}}=\pi\delta (x),
   \]
with
\[
  x\Rightarrow\omega ^{2}-\omega _{j}^{2}, \qquad
   \alpha\Rightarrow 2\omega\gamma_{j}(\omega)
    = 4\pi\omega L_j\sigma_{jj}(\omega).
     \]
Then, with accounting of Eq.(\ref{eq3}), the expression for the absorbed
energy takes the form
\ba
 w_{e}|_{\gamma \rightarrow 0}&=&{\frac{{V}}{32}}{\frac{{\omega _{p}^{2}}}{{
  \;\Gamma ^{2}}}}\sum\limits_{j=1}^{3}\Biggl\vert \mathrm{\,\mathbf{E}}_{0j}
   \left( \exp {\left[ {\ -{\frac{{(\omega _{j}-
    \omega_{0})^{2}}}{{4\Gamma ^{2}}}}}\right]}
     \right.\nonumber\\ &+& \left.
      \exp {\left[{\ -{\frac{{(\omega_{j}+\omega _{0})^{2}}}{{4\Gamma ^{2}}}}}
       \right] }\right)\Biggr\vert ^{2},
        \label{eq19}
         \ea
\noindent where
\begin{equation}
 \omega _{j}=\sqrt{L_{j}}\,\,\omega _{p}.
  \label{eq20}
   \end{equation}
We limit the analysis of Eq.~(\ref{eq19}) to the case of a
spheroidal MN and choose, for convenience, such a polarization of
the electric field, which makes it possible to excite the plasmon
oscillations of electrons both along and
across the spheroid rotation axis. If the field components $\mathrm{%
\mathbf{E}}_{0\parallel}$ and $\mathrm{\mathbf{E}}_{0\perp }$ are
directed along and across the rotation axis, respectively, then such
a polarization is, in the general case,
\begin{equation}
 E_{0\parallel} \equiv|\mathrm{\mathbf{E}}_{0}|\cos \theta ;
  \quad E_{0\perp }=|\mathrm{\mathbf{E}}_{0}|\sin \theta ,
   \label{eq21}
    \end{equation}
\vskip10pt
\noindent\includegraphics[width=8.6cm]{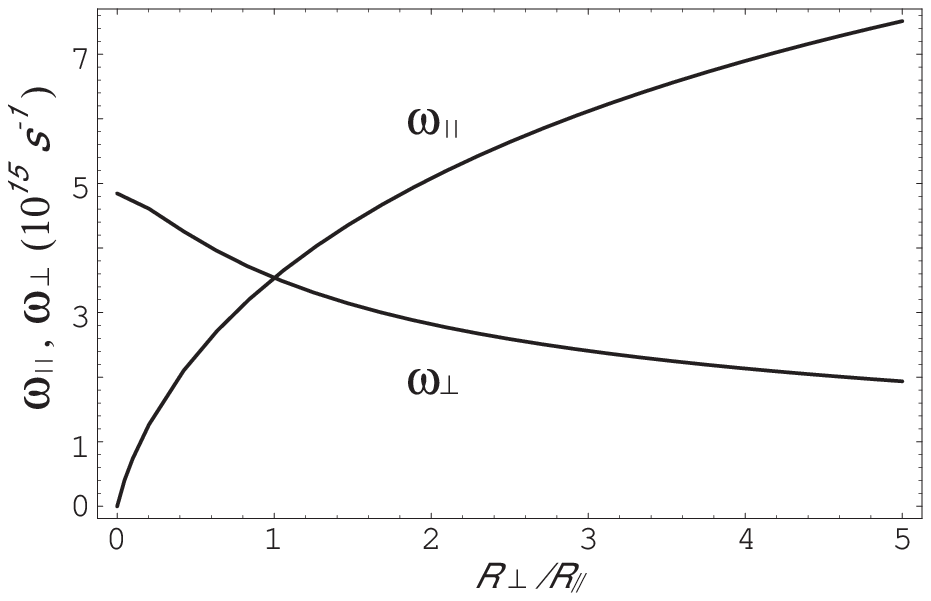}
\vskip-1mm\noindent{\footnotesize FIG. 1. Dependence of the plasmon
resonance frequencies on the form of gold nanoparticle.}
\vskip10pt
\noindent where $\theta $ is the angle between the spheroid rotation axis and
the vector of the incident electric field strength $\mathrm{\mathbf{E}}_{0}$.
Then, frequencies (\ref{eq20}) can be rewritten as the frequencies of
electron oscillations along and across the axis of rotation:
\begin{equation}
 \omega _{\parallel,\perp }=\sqrt{L_{\parallel,\perp }}\,\,\omega _{p},
  \label{eq22}
   \end{equation}
which, for the given $R_{\parallel,\perp }$ will be the frequencies
of the corresponding plasmon resonances. The dependence of the plasmon
resonance frequencies on the particle form is illustrated in Fig. 1 for Au.

We will use the electromagnetic radiation energy $w$ traversing the
nanoparticle during the total duration of the pulse as a scale. This
energy at normal light incidence, we define as

\begin{equation}
 w=s c\int_{-\infty}^{\infty}\mid {\bf E_0}(0,t)\mid^2 dt=
  \frac{s c}{2\Gamma}\sqrt{\frac{\pi}{2}}(1+e^{-\frac{
   \omega^2_0}{2\Gamma^2}})\mid {\bf E_0}\mid^2,
    \label{eq23}
     \end{equation}

\noindent where $c$ is the light velocity, and
\begin{equation}
 s=\pi(R^2_{\bot}R_{\parallel})^{2/3}
  \label{eq24}
   \end{equation}
is the particle section. In what follows, for the sake of illustration,
we will calculate the quantity
\begin{equation}
 S_{e}=\frac{w_e}{2w},
  \label{eq25}
   \end{equation}
which is the ratio of the energy absorbed by a unit of the MN volume
to the energy traversed the nanoparticle. The ratio between particle
volume and particle section is
\begin{equation}
 \frac{V}{s}=\frac{4}{3}R,
  \label{eq26}
   \end{equation}
\noindent where
\begin{equation}
 R=\sqrt[3]{R_{\bot}^2 R_{\parallel}}.
  \label{eq27}
   \end{equation}

\vskip10pt
\noindent\includegraphics[width=8.6cm]{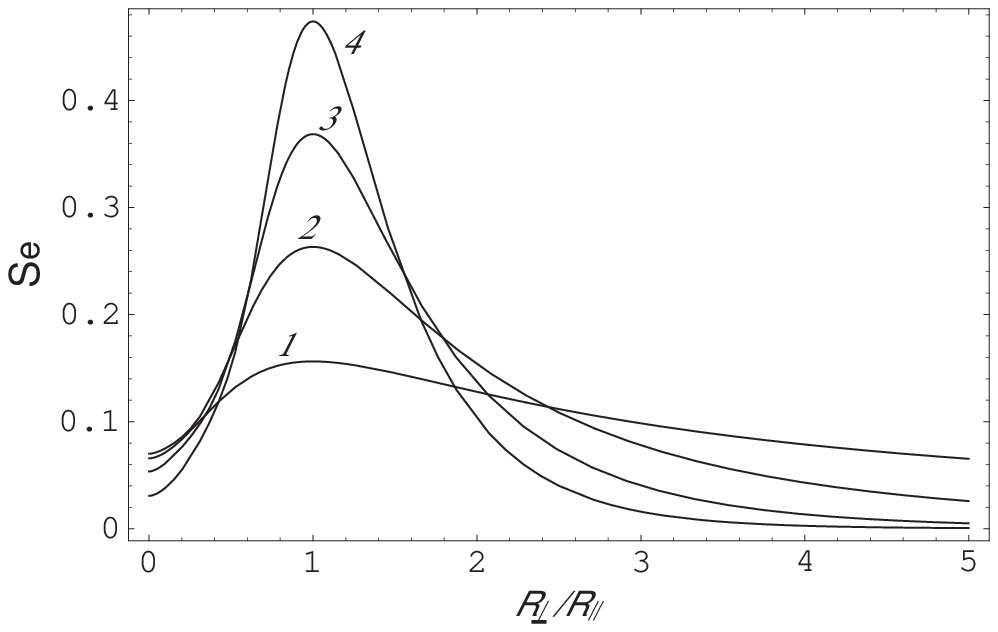}
\vskip-1mm\noindent{\footnotesize FIG. 2. Dependence of the energy
absorbed by a spheroidal gold particle with
$R=(R_{\bot}^2 R_{\parallel})^{1/3}=200 \AA$ at the plasmon resonance
frequency on the ratio between the spheroid semiaxes for different
values of $\Gamma $, s$^{-1}$: $2.637\times 10^{15}$ (\textit{1%
}), $1.582\times 10^{15}$ (\textit{2}), $1.13\times 10^{15}$
(\textit{3}), and $0.879\times 10^{15}$ (\textit{4})}. $\theta =\pi/4$.
\vskip10pt

Figure 2 shows the dependence of $S_{e}$ on the degree of
ellipsoid's oblateness or prolateness in the case where $\theta =\pi
/4$ and the frequency equals that of surface plasmons of a spherical
particle $\omega _{0}=\omega _{p}/\sqrt{3}\equiv \Omega $. Here and
below, the calculations are carried out with the use of formulas
(\ref{eq19}) and (\ref{eq21}) -- (\ref{eq27}), given electron density
in the Au particle\cite{CK} $n\approx5.9\times$10$^{22}$ cm$^{-3}$,
$\omega _{p}=(4\pi n e^2/m)^{1/2}\approx 1.37\times 10^{16}$ s$^{-1}$
and $\Omega\approx7.91\times 10^{15}$ s$^{-1}$. The numerical data
for $\Gamma$ are taken here and below (in illustrations) from the
Table 1. It is convenient to measure $\Gamma$ in the fractions
of the carrier frequency $\omega _{0}$. Then, the duration of a laser
pulse can be set in the carrier wave periods or wavelengths, that is
$1/\Gamma_{n}=n/\omega_0\sim{n \lambda_0}$, where $n$ is an arbitrary
integer. Thus, for the region of plasmon resonances (when
$\omega _{0}\equiv\Omega $), each wavelength, which is a multiple
of $\lambda _{0}$, can be brought into correspondence with
a certain pulse duration.

\vskip3mm
\noindent{\footnotesize{\bf
TABLE I. Correspondence between the value inverse to the pulse
duration and the wavelength of a carrier wave in the region of
plasmon resonances for Au particle in vacuum.}
\vskip1mm \tabcolsep42.9pt
 \noindent
  \begin{tabular}{c c}
   \hline
    \hline
     \multicolumn{1}{c}{\rule{0pt}{9pt}$\Gamma \times 10^{15}$, s$^{-1}$} &
      \multicolumn{1}{c}{$\lambda$}\\%
       \hline%
        \rule{0pt}{9pt} 7.911 & $\lambda _{0}$ \\
         2.637 & 3$\lambda _{0}$ \\
          1.582 & 5$\lambda _{0}$ \\
           1.130 & 7$\lambda_{0}$ \\
            0.879& 9$\lambda _{0}$ \\
             0.791 &10$\lambda _{0}$ \\
              0.527&15$\lambda _{0}$ \\
               0.396 &20$\lambda _{0}$ \\
                \hline
                 \hline
                  \end{tabular}}
                   \vskip10pt
As follows from curves \textit{1}--\textit{4}, the laser
pulses of longer duration are absorbed more effectively.
As for the shape dependence of the absorption, the maximum
of absorption is achieved at $R_{\perp}/R_{\parallel }=1$,
i.e., when the particle is in vacuum and have the spherical
shape. If the particle is embedded in the dielectric matrix
with $\varepsilon >1$, then maximum is shifted to the red side
of  the spectra range, proportionally to the value of $\varepsilon$.
For example, for Au particle embedded in the glass matrix with
$\varepsilon=7$, the plasmon resonance pick occurs at the frequency
$\Omega_m\simeq\Omega - 0.55\Omega$.

The calculation of $S_{e}$ was also carried out by means of the
numerical integration in formula (\ref{eq16}) for the various
particle radii $R$ within the range of
$R=(R_{\parallel }R_{\perp }^{2})^{1/3}=50\div 400 $~{\AA}.
On the whole, it turned
out that the value of $S_{e}$, calculated numerically, is (i)
slightly smaller than the values obtained on the basis of formula
(\ref{eq19}) and (ii) depends on $R$ very weakly for
various~\cite{foo1} $\Gamma$.
At the same time, the latter dependence is more
pronounced at smaller $\Gamma $ provided that $R$ is not great
($R=50\div 150$ {\AA}). For example, for particle with $R=100$
{\AA} at $\Gamma =0.879\times 10^{15}$ s$^{-1}$, the highest possible
values of $S_{e}$ calculated analytically with the use of
Eq.~(\ref{eq19}) exceed those numerically determined from formula
(\ref{eq16}) by about $7\%$. For greater $R$ and $\Gamma$, this
percentage decreases. For this reason, we will rest in the further
calculations of $S_{e}$ on the evaluations performed according to
formula~(\ref{eq19}).\looseness=1

\vskip10pt
\noindent\includegraphics[width=8.6cm]{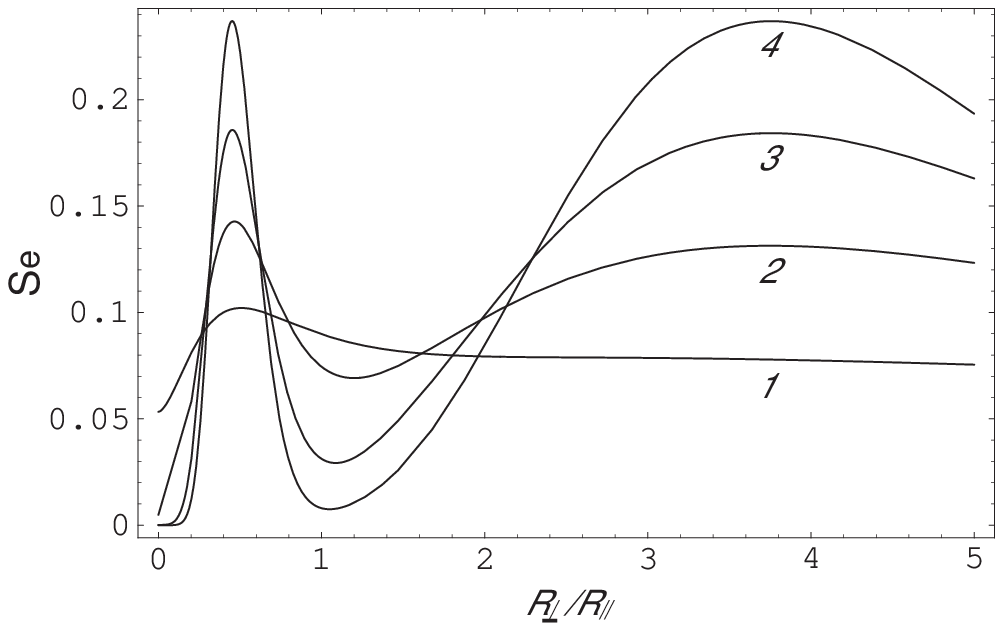}
\vskip-1mm\noindent{\footnotesize FIG. 3. The same as in Fig. 1,
for the frequency $\omega_{0}=5.4\times 10^{15}$~s$^{-1}$}.
\vskip10pt

As soon as the carrier frequency deviates from that of the
surface plasmon in a spherical particle, the peak of absorbed energy
depending on the ratio $R_{\bot }/R_{\parallel }$ splits into the
two ones: the first maximum corresponds to the prolate particles
($R_{\perp}/R_{\parallel }<1$), whereas the second -- to the oblate
particles ($R_{\bot }/R_{\parallel }>1$). This fact is illustrated
in Fig. 3 for the frequency $\omega _{0}=5.4\times 10^{15}$
s$^{-1}$ and various values of the incident pulse
duration.

If the pulse duration is fixed, then, changing the frequency of a
carrier wave, we can trace the evolution of the dependence of the
maximum of absorbed energy on the MN shape. Figures 4(\textit{a}) and
4(\textit{b}) illustrate such a dependencies. For comparison, the
absorption curve at a frequency $\omega _{0}=\Omega $, where $\Omega
$ is the resonance frequency of a spherical MN, is shown (curve
\textit{1} in each part of the figure). For the frequencies
exceeding $\Omega $ (see Fig. 4(\textit{a})), the absorption maximum
first shifts towards the side of more {\it oblate} particles with increase
in $\omega _{0}$ and then splits into the two peaks: the first one is
observed at $R_{\perp}/R_{\parallel }>1$, whereas the second one --
at $R_{\perp }/R_{\parallel }<1$. Figure 4(\textit{b}) demonstrates
the data calculated for the frequencies lower than $\Omega $, with
the calculations carried out for the same values of the
pulse duration. As the frequency gets lower, the absorption maximum
shifts towards the side of more {\it prolate} particles and, afterwards,
splits into two, as in the former case. As $\omega _{0}$ deviates
from the resonance frequency, the peak of the absorbed energy first
decreases in absolute value, then splits and finally, as the carrier
frequency strongly differs from $\Omega $, the values of both
emerged peaks reach the plateau.\looseness=1

\vskip10pt
\noindent\includegraphics[width=8.6cm]{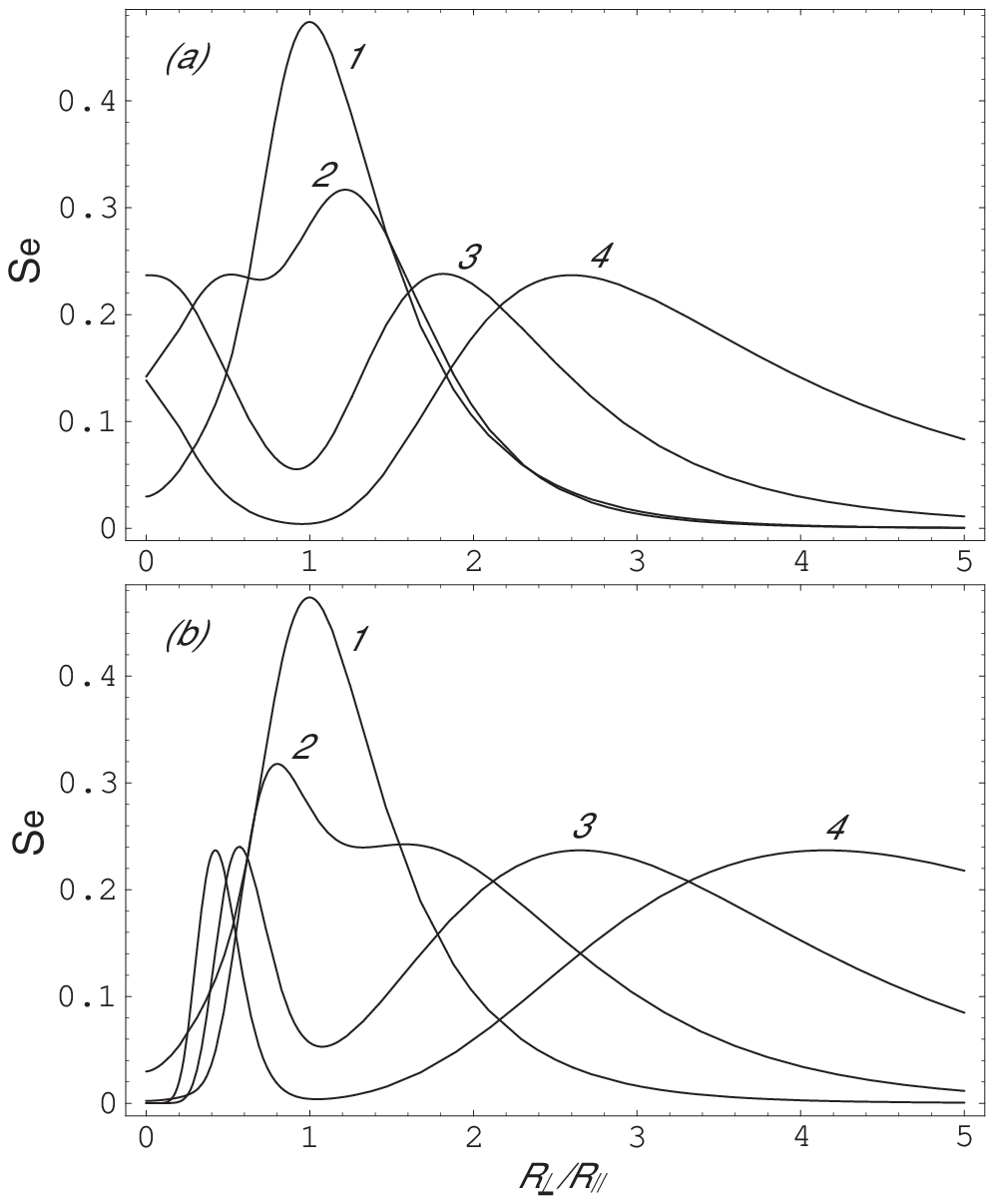}

\vskip-1mm\noindent{\footnotesize FIG. 4. The energy absorbed by a
spheroidal gold particle with $R=(R_{\bot}^2 R_{\parallel})^{1/3}=200 \AA$
 as the function of the degree of its oblateness or
prolateness for $\Gamma=0.879\times 10^{15}$ s$^{-1}$, at
the frequencies $\omega _{0}$ higher than\ that of the plasmon
resonance $\Omega $ of a spherical MN (\textit{a}): $\omega
_{0}=8.8\times 10^{15}$ s$^{-1}$ (\textit{2}), $9.7\times
10^{15}$ s$^{-1}$ (\textit{3}), and $10.6\times 10^{15}$ s$^{-1}$
(\textit{ 4}) and lower than\ $\Omega $ (\textit{b}): $\omega
_{0}=7.0\times 10^{15}$
 s$^{-1}$ (\textit{2}), $6.1\times 10^{15}$ s$^{-1}$ (\textit{3}%
), and $5.2\times 10^{15}$ s$^{-1}$ (\textit{4}). $\theta =\pi/4$.
Curves \textit{1} in both parts of the figure correspond
to the case $\omega _{0}=\Omega $.}
\vskip10pt
\noindent These results are in compliance with the correlation, established
by us earlier,\cite{GT4} between the geometrical shape of a nanoparticle
and the frequency of the light absorption, depending on them, the rapid
growth in the strength of light pressure on a particle is observed.

Let us choose the prolateness or oblateness of a nanoparticle fixed
and trace how the absorption changes as a function of the carrier
frequency. In the case of a {\it prolate} MN, with increase in the
duration of the incident laser pulse, one observes not only the
growth of the peak of absorbed energy, but also its splitting
(as $\Gamma $ becomes higher than a certain critical value) into
two peaks (a doublet) located on the opposite sides of the line
$\omega _{0}/\Omega =1$ (Fig. 5), where $\Omega $ corresponds to
the resonance frequency of a spherical particle. It is worth to note
that the peaks at $\omega _{0}<\Omega $ and $\omega _{0}> \Omega $
correspond to the plasmon resonance frequencies $\omega _{\parallel }$
and $\omega _{\perp }$, respectively (see Eq.~(\ref{eq22}) and Fig. 1),
whereas the inverse situation is characteristic of {\it oblate}
particles: the former peak corresponds to $\omega _{\perp }$
and the latter -- to $\omega _{\parallel }$ (see Fig. 1).

\vskip10pt
\noindent\includegraphics[width=8.6cm]{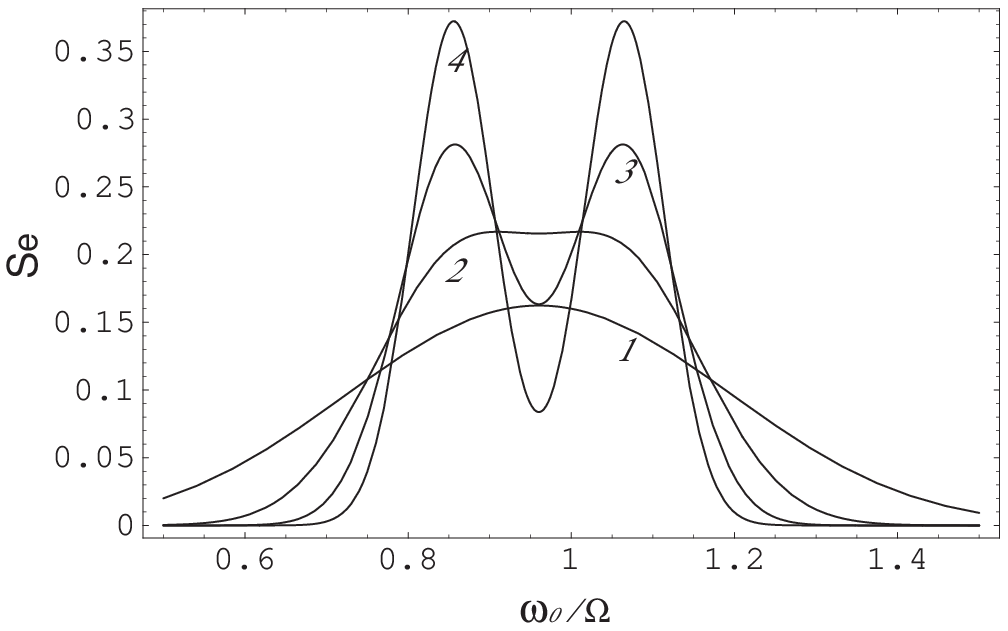}

\vskip-1mm\noindent{\footnotesize FIG. 5. Dependence of the energy
absorbed by a prolate ($R_{\bot }/R_{\parallel }=0.7$) gold
particle with $R=(R_{\bot}^2 R_{\parallel})^{1/3}=200 \AA$
on the frequency of a carrier wave for different values of
$\Gamma $, s$^{-1}$: $1.582\times 10^{15}$ (\textit{1%
}), $0.791\times 10^{15}$ (\textit{2}), $0.527\times 10^{15}$
(\textit{3}), and $0.396\times 10^{15}$ (\textit{4})
(taken from Table 1). $\theta =\pi/4$.}%
\vskip10pt

\noindent

Thus, contrary to the case of spherical particles, the
spheroidal MNs, either prolate or oblate, exhibit a doublet
in the spectrum of absorption provided that $\Gamma $ is
sufficiently small. The doublet originates from two plasmon
resonances excited in particles by ultrashort laser pulses.
At $\theta =\pi /4$, both the peaks of the doublet have the same height,
which, being practically independent of a degree of the particle
prolateness or oblateness, strongly depends on the pulse duration.
In the case of prolate MNs, a minimum of the dip between the absorption
peaks is achieved at frequencies $\omega_0$ lower than $\Omega $
and shifts towards the longer waves with increase in the prolateness
degree. The other tendency is characteristic of oblate MNs: the
minimum is observed at frequencies higher than $\Omega $ and, as the
oblateness degree increases, it first shifts towards the shorter
waves and then, at sufficiently high oblateness degrees, the
movement changes its direction.\looseness=1

The peaks of absorption display the similar behavior. As the degree
of the MN prolateness increases, the distance between the peaks of the
doublet grows due to the more drastic shift of the peak, caused by
the plasmon resonance at the frequency $\omega_\parallel$, toward
the longer waves. At the same time, the height of the peaks remains
constant provided that the degree of prolateness is small enough.
Quite similar behavior is also characteristic of oblate MNs: the
distance between the peaks of the doublet grows with increase in the
degree of the MN oblateness. The reason for this is either the more
drastic shift of the peak, caused by the plasmon resonance at
$\omega _{\parallel }$, toward the highfrequency side of the spectrum
(at sufficiently small degrees of oblateness) or the reverse
movement of the peak, caused by the plasmon resonance at $\omega_{\perp }$
(at higher degrees of oblateness). The height of the peaks remains
constant as the degree of oblateness keeps small enough.

Under various experimental conditions, a small shift, either blue
or red, with respect to the value $\omega _{0}=\Omega $ has been
observed. A number of microscopic approaches has been developed to
explain such a shift (see, for example, Ref. \onlinecite{Hee}). The
most successful approaches are thought to be those built on the basis
of a jelly model with the use of the theory of linear response in
the limits of time-dependent local density.\cite{Eka} However, these
approaches don't account for the shift caused by a change in the
particle shape. It is the most important feature of our approach
that results in the facts that the shape of the MNs specifies the
frequency, at which the resonance absorption occurs, and the shift
of the plasmon absorption peak is associated with a change in the
particle shape.

\vskip10pt
\noindent\includegraphics[width=8.6cm]{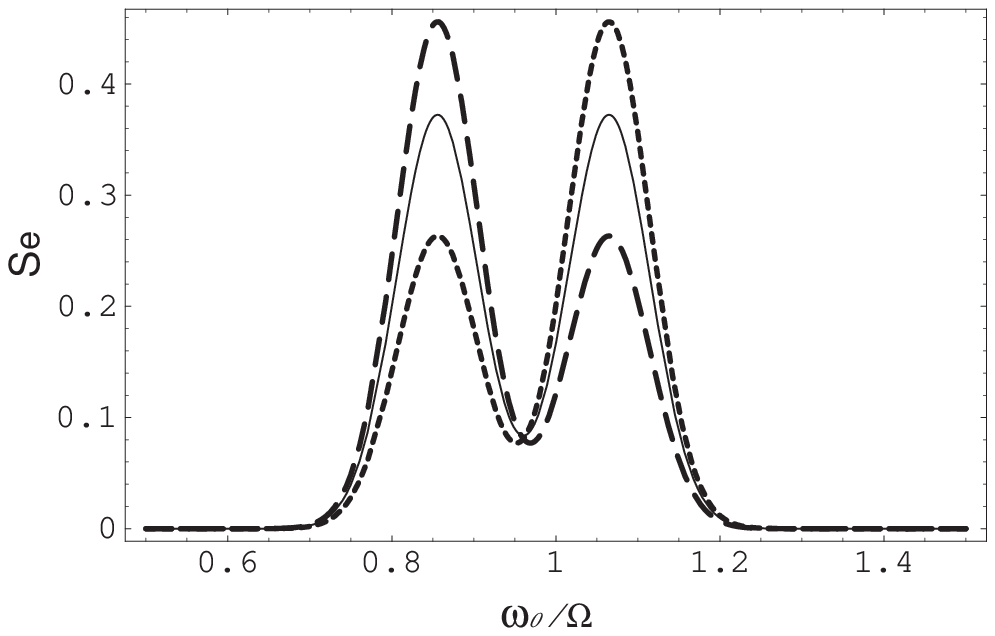}

\vskip-1mm\noindent{\footnotesize FIG. 6. The energy absorbed
by a prolate ($R_{\bot }/R_{\parallel }=0.7$) gold particle
with $R=(R_{\bot}^2 R_{\parallel})^{1/3}=200 \AA$ as the
function of the frequency of a carrier wave for
$\Gamma =0.396\times 10^{15}$, s$^{-1}$ and different angles of the
laser pulse incidence: $\theta =\pi /4$ -- solid line, $\theta
=\pi /3$ -- dotted line, and $\theta =\pi /6$ -- dashed line.}
\vskip10pt

The relative height of the peaks in the doublet can be tuned by
means of changing the angle of the laser pulse incidence. Figure 6
allows one to trace how the plasmon absorption along and across the
spheroid rotation axis changes as the angle of incidence $\theta $
is reduced from $\pi /3$ to $\pi /6$. For convenience, we consider
the prolate particle with $R_{\bot }/R_{\parallel }=0.7$, as in
Fig. 5. As is seen in Fig. 6, the peak intensities are the same
at $\theta =\pi /4$, but they become different as $\theta $ deviates
from this value. So, for example, at $\theta =\pi/3$, the peak
associated with the resonance at $\omega _{\parallel}$ is
characterized by a reduced intensity, contrary to the peak associated
with $\omega _{\perp }$, which is characterized by an increased intensity.
At $\theta =\pi /6$, however, the inverse situation is observed. It is
obviously that, whereas the intensity of the peak at $\omega
_{\parallel}$ reaches a maximum at $\theta =0$ and turns into the zero
at $\theta =\pi /2$, the peak at $\omega _{\perp}$ has maximum
intensity at $\theta =\pi /2$ and minimum one -- at $\theta =0$.

Lastly, we consider the qualitatively different limit transition
when the width of an incident laser pulse is sufficiently large.
To do this, we use the next representation for $\delta $-function\cite{Arf}
\[
 \delta (x)={\frac{{1}}{\sqrt{\pi }}}\lim_{\alpha \rightarrow 0}{\frac{{1}}{{
  \alpha }}}\exp (-x^{2}/\alpha ^{2}),
   \]
and calculate, according to Eq.~(\ref{eq3}), the value of
$|{E^{(0)}_j}(0,\omega)|^2$, which is included in Eq.~(\ref{eq16})
\begin{equation}
 \left|E^{(0)}_j(0,\omega )\right| ^{2}
  \simeq 2{E}_{{0j}}^{{2}}\left( {{\frac{{\pi }}{{2}}}}%
   \right) ^{3/2}{\frac{{1}}{{\Gamma }}}{\left[ {\delta (\omega
   -\omega _{0})+\delta (\omega +\omega _{0})}\right] }.
     \label{eq28}
      \end{equation}
Now relations (\ref{eq16}) and (\ref{eq28}) yield
\begin{equation}
 w_{e}|_{\Gamma \rightarrow 0}={\frac{{2V}}{{\pi }}}
  \left( {\frac{{\pi }}{2}}\right) ^{3/2}{\frac{{1}}{{\Gamma}}}{\sum
   \limits_{j}{\frac{{\omega _{0}^{4}\;\sigma _{jj}(\omega_{0})
    \;E_{0j}^{2}}}{{(\omega _{0}^{2}-\omega _{j}^{2})^{2}+(2\gamma_j(\omega _{0}))^{2}
     \omega _{0}^{2}}}}}.
      \label{eq29}
       \end{equation}
As is seen from Eq.~(\ref{eq29}), the absorbed energy comparing
with Eq. (\ref{eq19}) is proportional only to the first order
of the pulse duration $1/\Gamma $. Expression
(\ref{eq29})\ divided by the pulse duration gives us the value of
the energy which is absorbed by a nanoparticle on average in a unit
of time at its irradiation by long-duration pulses. This quantity
differs by the factor $2\sqrt{\pi /2}$ from the
analogous quantity obtained for the absorption of a spherical
wave.\cite{foo2}

The energy absorbed
by spheroidal MP in the case of broad pulses is given
from Eqs. (\ref{eq29}), (\ref{eq18}) by

\begin{widetext}
 \begin{eqnarray}
 w_{e}|_{\Gamma \rightarrow 0}=\frac{9}{32}\frac{|{\bf E}_0|^2}{\sqrt{2\pi}}
  \frac{V}{\Gamma}\omega^2_0\omega^2_p\frac{v_F}{R_{\bot}}
   {\left({\frac{\eta(e_s) \cos^2\theta}{{(\omega _{0}^{2}-
    \omega _{\parallel}^{2})^{2}+[2\gamma_{\parallel}(\omega _{0})]^{2}
     \omega _{0}^{2}}}}+{\frac{\rho(e_s) \sin^2\theta}{{(\omega _{0}^{2}-
      \omega _{\bot}^{2})^{2}+[2\gamma_{\bot}(\omega _{0})]^{2}
       \omega _{0}^{2}}}}\right)},
        \label{eq30}
         \end{eqnarray}
          \end{widetext}
where
\begin{equation}
 \gamma_{\left(\parallel \atop \bot\right)}(\omega_0)=2\pi L_{\left(\parallel
  \atop \bot\right)} \sigma_{\left(\parallel \atop \bot\right)}(\omega_0),
   \label{eq31}
    \end{equation}
and $L_{\parallel}$, $L_{\bot}$ are the geometrical factors along
and across the long spheroid axis of a revolution, correspondingly.

\section{MAGNETIC ABSORPTION}

If the frequency of a monochromatic EM wave is far from a plasmon resonance,
then a separate mechanism plays the main role in the MN absorption.
As we have already pointed out in this case, either an electric or magnetic
absorption can dominate. But if the size of a spherical particle, for instance,
is larger than 50 \AA, then the magnetic absorption begin to exceed the electric
one even at frequencies less than the frequency of electron reflections from
opposite walls of a particle.\cite{TG} In this section, we will consider in detail,
the behavior of the magnetic absorption at frequencies close to the plasmon
resonances of the MN.

In accordance with Eqs.~(\ref{eq11}) and (\ref{eq12}), the magnetic
field energy can be presented in an integral form as follows

\begin{equation}
 w_{m} = {\frac{{1}}{{2}}}{\int_{ - \infty} ^{\infty}
  {{\frac{{d\omega }}{{2\pi}} }\;{\int_{V} {d{\rm {\bf r}}}}} }
   \;[{\rm {\bf j}}_{ed} ({\rm {\bf r}},\omega )
    \;{\rm {\bf E}}_{ed}^{\ast}  ({\rm {\bf r}},\omega )+c.c.].
     \label{eq32}
      \end{equation}

\noindent The value of ${\rm {\bf E}}_{ed} ({\rm {\bf r}},\omega )$
is already known for us from Eq.~(\ref{eq9}). In order to calculate
${\rm {\bf j}}_{ed} ({\rm {\bf r}},\omega )$ with the use of the
formula (\ref{eq13}), it is necessary first to find the distribution
function for electrons. For this purpose, we would seek for solutions
of Eq.~(\ref{eq14}), where $E_{in}(\omega)=0$. To satisfy both the
equation (\ref{eq14}) and the boundary conditions (\ref{eq15}),
we should pass to the deformed variables

\[
 x_{i} ' = x_{i} {\frac{{R}}{{R_{i}}} },
  \quad
   \upsilon _{j} ' = \upsilon _{j} {\frac{{R}}{{R_{j}}} }.
    \]

\noindent Then an ellipsoid can be represented as a sphere
of the same volume with the radius $R$, and the solution of
Eq. (\ref{eq14}) we can write in the form
\ba
 f_{1} ({\rm {\bf r}}',{\rm {\bf v}}',\omega ) &=& -{\frac{{e}}{{R^{2}}}}{
  \frac{{\partial f_{0}}} {{\partial \varepsilon}}}\sum\limits_{i,j = 1}^{3}
   \alpha _{ij} (\omega )\;\upsilon'_{j}\, R_{j}\, R_{i}
    \nonumber\\
    &\times &
      \left( {x'_{i} + \upsilon '_{i} {\frac{{\partial}}{{\partial (\nu - i\omega )}}}}
       \right) {\frac{{1-e^{ - (\nu - i\omega )\,t'({\rm {\bf r}}',{\rm {\bf v}}')}}}{{
        \nu - i\omega}} },
         \nonumber\\
          \label{eq33}
           \ea
\noindent where

\begin{equation}
 t'({\rm {\bf r}}',{\rm {\bf v}}') =
  {\frac{{1}}{{\upsilon '^{2}}}}[{\rm {\bf r}}'{\rm {\bf v}}' +
    \sqrt{(R^{2} - r'^{2})\upsilon '^{2} + ({\rm {\bf r}}'{\rm {\bf v}}')^{2}}].
     \label{eq34}
      \end{equation}

\noindent In Eq.~(\ref{eq33}), the energy derivative of the
distribution function in zero approximation in a small ratio of
${{k_{B} T} \mathord{\left/ {\vphantom {{k_{B} T} {\varepsilon_{F}}} }
\right. \kern-\nulldelimiterspace} {\varepsilon _{F}}} $
 can be replaced by

\begin{equation}
 {\frac{{\partial f_{0}}} {{\partial \varepsilon}} } \to -
   \delta (\varepsilon - \varepsilon _{F} ),
    \label{eq35}
     \end{equation}
where $\varepsilon_F$ it the Fermi energy.
\noindent The
diagonal components of the matrix $\alpha _{jj} $ are $\alpha _{jj} $=0.
The nondiagonal ones can be expressed in terms of appropriate
components of the magnetic field. For example,

\begin{equation}
 \alpha _{xy} (\omega ) = - i{\frac{{\omega}}{{c}}}{
  \frac{{R_{x}^{2} }}{{R_{x}^{2} + R_{y}^{2}}} }H_{z}^{(0)}(0,\omega ).
   \label{eq36}
    \end{equation}

\noindent The other two components of the matrix $\alpha $ can be obtained
from Eq.~(\ref{eq36}) by means of a cyclic index permutation. The remaining
three components one can find using the antisymmetric character of the
$\alpha $, namely, taking into account the property $\alpha _{xy}
(\omega ) = - \alpha _{yx} (\omega )$.

Thus, to find the energy of the magnetic absorption it is necessary
to undertake step-by-step the following procedure: to inset the
obtained function $f_{1} ({\rm {\bf r}},{\rm {\bf v}},t)$
into Eq.~(\ref{eq13}) and then, taking into account the
expression (\ref{eq9}), to put the Fourier components for both
the current density and the curly field into Eq.~(\ref{eq32}).

One can escape the above procedure for energy calculation,
using the phenomenological approach for an estimation of current.
It means that the absorption of a particle can be written down in a
general form with real and imaginary parts of a dielectric permeability,
and thus it is not necessity to solve the kinetic equation (\ref {eq14})
with the boundary conditions (\ref{eq15}).

\subsection{Phenomenological approach}

Assume that the particle size is much more greater than the electron
free pass within it. Then, the curly current can be expressed in terms
of a curly field by means of

\[
\rm {\bf j}_{ed} (\omega ) = \sigma _{m} (\omega)\;{\rm {\bf E}}_{ed} (\omega ),
\]

\noindent and the expression (\ref{eq32}) can be rewritten in the form

\begin{equation}
  w_{m} = {\frac{{1}}{{2}}}{\sum\limits_{j = 1}^{3}{{\int_{ -
   \infty }^{\infty}  {[\sigma _{m} (\omega)+\sigma^*_{m} (\omega)]\;{\frac{{d\omega}} {{2\pi }}}
    \;{\int_{V} {\vert {\rm {\bf E}}_{ed}^{j} ({\rm {\bf r}},
     \omega )\vert ^{2}\;d{\rm {\bf r}}}}} }}} .
      \label{eq37}
       \end{equation}

\noindent One can change here the integration over an ellipsoid volume $V$
by the integration over the sphere of an equivalent volume.
Then, it is easy to show, that

 \begin{eqnarray}
&&{\int_{V} {x^{2}}} d{\rm {\bf r}} = V{\frac{{R_{x}^{2}}} {5}},
    \nonumber \\
  &&{\int_{V} {y^{2}}} d{\rm {\bf r}}
   =V{\frac{{R_{y}^{2}}}{5}}, \quad {\int_{V} {z^{2}}} d{\rm {\bf r}}=
     V{\frac{{R_{z}^{2}}} {5}},
        \nonumber \\
      &&{\int_{V} {xy\;}} d{\rm {\bf r}} =
         {\int_{V} {yz\;}} d{\rm{\bf r}} = {\int_{V} {zx\;}} d{\rm {\bf r}} = 0.
           \label{eq38}
            \end{eqnarray}

Further, we will restrict ourselves with consideration of the metal
nanoparticles having the spheroid shape ($R_{x} = R_{y} \equiv R_{
\bot} $, $R_{z} \equiv R_{\vert \,\vert} )$. Using Eqs.~(\ref{eq9})
and (\ref{eq38}), the sum of integrals over a nanoparticle volume will be

\ba
 &&{\sum\limits_{j = 1}^{3} {{\int_{V} {\vert {\rm {\bf E}}_{ed}^{j} ({\rm {\bf r}},\omega )
  \vert ^{2}d{\rm {\bf r}}}}} }=
   \frac{V}{5}\left( {{\frac{{\omega}}{c}}}\right )^{2}
    \nonumber\\
    &\times &
      \left( {{\frac{{R_{ \bot} ^{2} }}{2}}\;{\rm {\bf H}}_{\,\vert
       \vert} ^{{\rm {\bf 2}}} (0,\omega ) + {\frac{{R_{\vert
        \,\vert} ^{2} R_{ \bot} ^{2}}} {{R_{\vert \,\vert} ^{2} + R_{ \bot}^{2}}} }
         \;{\rm {\bf H}}_{ \bot} ^{{\rm {\bf 2}}} (0,\omega )}
          \right).
           \label{eq39}
            \ea

Here, ${\rm {\bf H}}_{\,\vert \vert} ^{} (0,\omega )$ is the
intensity of the magnetic field along the spheroid axis of a revolution,
and ${\rm {\bf H}}_{ \bot}(0,\omega )$ -- transverse to it. It is worth
to pay attention that the magnetic absorption depends on the magnetic
field polarization in the same way as the electric absorption depends
on the electric field polarization.\cite{LL} Moreover, one can generalize
the expression (\ref{eq39}) to the case of an arbitrary coordinate
systems if the field components are represented as

\begin{equation}
 {\rm {\bf H}}_{ \bot} ^{2} = {\rm {\bf H}}^{2} - {\rm
  {\bf H}}_{\vert \,\vert} ^{2} , \qquad {\rm {\bf H}}_{\vert \,\vert}
= ({\rm {\bf H}}\cdot {\rm {\bf n}})\,{\rm {\bf n}},
     \label{eq40}
      \end{equation}

\noindent where ${\rm {\bf n}}$ is a unit vector directed along the
spheroid axis of a revolution. Then with an account of Eq.~(\ref{eq39}),
the expression (\ref{eq37}) transforms into
\ba
 w_{m} &=& {\frac{{V}}{{10}}}{\frac{{R_{\vert \,
  \vert}^{2} R_{ \bot} ^{2} }}{{R_{\vert \,\vert} ^{2} + R_{ \bot} ^{2}}}}
   \int_{ - \infty} ^{\infty} [\sigma _{m} (\omega )+\sigma^*_{m} (\omega)]\,
    \left({{\frac{{\omega}} {{c}}}} \right)^{2}
     \nonumber\\
     &\times &
       \left( {{\rm {\bf H}}^{2}(0,\omega ) + {\frac{{1}}{{2}}}
        \left( {1 - {\frac{{R_{ \bot}^{2}}} {{R_{\vert \,\vert} ^{2}}} }}
         \right)\;({\rm {\bf H}}(0,\omega )\,{\rm {\bf n}})^{2}}
          \right){\frac{{d\omega}}{{2\pi}} } .
           \nonumber\\
            \label{eq41}
             \ea

\noindent The field ${\rm {\bf H}}(0,\omega )$ in Eq.~(\ref{eq41})
is defined by the expression (\ref{eq4}), and $\sigma _{m}$ for
a metallic nanoparticle in the phenomenological case is

\begin{equation}
  \sigma _{m} (\omega ) = {\frac{{1}} {{4\pi}}}{
   \frac{{\omega_{p}^{2} }}{{\nu -i \omega}}},
    \label{eq42}
     \end{equation}

\noindent where $\omega_{p} = 4\pi \,n\,e^{2} / m$
is the plasma frequency of the electron vibrations in a metal.

Next, we will perform calculations for certain polarizations
of the magnetic component of the EM wave with respect to the particle
orientation. We will consider the following two cases:

\smallskip

(i) Let us direct the vector of magnetic field transverse to the
spheroid axis. Then, the second item under the integral in Eq.~(\ref{eq41})
vanishes, and after substituting Eqs.~(\ref{eq4}) and (\ref{eq42}) into
Eq.~(\ref{eq41}), we get for the absorbed energy the expression

\ba
 w_{m \bot}  &=& {\frac{{V}}{{80\;\Gamma^{2}}}}{\frac{{\nu
  \omega _{p}^{2} }}{{c^{2}}}}{\frac{{R_{\vert
   \,\vert} ^{2} R_{ \bot} ^{2}}} {{R_{\vert
    \,\vert} ^{2} + R_{ \bot}^{2}}} }\vert {\rm {\bf E}}_{0} \vert ^{2}
     \nonumber\\
     &\times &
    {\int_{ - \infty}^{\infty} {{\frac{{\omega ^{2}}}{{\nu ^{2} +
        \omega^{2}}}}\;f(\omega )\;{\frac{{d\omega}} {{2\pi}} }}} ,
         \label{eq43}
          \ea
\noindent where

\begin{equation}
f(\omega ) = \left( {\exp {\left[ { - {\frac{{(\omega -
   \omega _{0} )^{2}}}{{4\Gamma ^{2}}}}} \right]} + \exp {\left[ { -
   {\frac{{(\omega + \omega _{0} )^{2}}}{{4\Gamma ^{2}}}}} \right]}}
     \right)^{2}.
      \label{eq44}
       \end{equation}

\noindent Our attempt to fulfil the integration in Eq.~(\ref{eq43}) in
an analytical form exactly was not successful, but using the fact that
the frequency of electron collisions with phonons $\nu $ is a small
quantity comparing to the frequencies $\omega \approx \omega_{0}$,
which yield the main contribution to the integral, one can neglect
by $\nu$ under the integral in Eq.~(\ref{eq43}). For instance,
$\omega_{0} \sim $10$^{15}$ s$^{{\rm -}1}$ and for Au
$\nu_{0^o {\rm C}}\simeq $3.39$\times 10^{13 }$ s$^{{\rm -}1}$,
that comes from the formula for electroconductivity of metal
$\sigma=1/\rho\simeq ne^2/(m\nu)$, with
$\rho_{0^o {\rm C}}=2.04\times 10^{-6} \Omega\;{\rm cm}$ (Ref. \onlinecite{KL}).
Then, one gets the result equal to twice the integral of

\begin{equation}
 I_{\nu}  = {\int\limits_{\nu} ^{\infty}  {f(\omega)\;}} d\omega .
  \label{eq45}
   \end{equation}

\noindent This integral can be calculated analytically even in the
pointed limits. However, it is easy to show, that the result of the
integration practically is not changed if one brings the lower
integration limit in Eq.~(\ref{eq45}) down to the zero. This means
only that the integral between limits [0,$\;\nu $]

\begin{equation}
 I = {\int\limits_{0}^{\nu}  {f(\omega )\;}} d\omega \ll I_{\nu}
  \label{eq46}
   \end{equation}

\noindent is much less than one of $I_{\nu}$, and one can neglect
by it. Thus, for the integral between limits $[0,\infty ]$, we have:

 \ba
&&\int\limits_{0}^{\infty} {\left\{ {\exp {\left[ { -{\frac{{(\omega -
   \omega _{0} )^{2}}}{{4\Gamma ^{2}}}}} \right]} +
    \exp {\left[ { - {\frac{{(\omega + \omega _{0} )^{2}}}{{4\Gamma^{2}}}}}
     \right]}} \right\}^{2}\;} d\omega
      \nonumber\\
      & = &
        \sqrt {2\pi}  \;\Gamma
         \left( {1 + \exp {\left[ { - {\frac{{\omega _{0}^{2}}} {{2\Gamma^{2}}}}}
          \right]}} \right),
           \label{eq47}
            \ea
\noindent and as a result

\ba
 w_{m \bot}  &=& {\frac{{1}}{{40}}}{\frac{{V}}{{\sqrt{2\pi}} } }{
  \frac{{\nu }}{{\Gamma}} } \;\left( {{\frac{{\omega _{p}R_{ \bot}} } {{c}}}}
   \right)^{2}\left( {{\frac{{R_{\vert \,\vert}^{2}}} {{R_{\vert \,
    \vert} ^{2} + R_{ \bot} ^{2}}} }}\right)\,
     \nonumber\\
    & \times &
       \left( {1 + e^{ - {\frac{{\omega _{0}^{2} }}{{2\Gamma^{2}}}}}}
        \right)\vert {\rm {\bf E}}_{0} \vert ^{2},
         \label{eq48}
          \ea
\noindent provided that $\omega \gg \nu $.

\smallskip

(ii) Now we pass to the case of EM wave-polarization, where
the magnetic field is directed along the main spheroid axis. In this
case the scalar product in a second item under the integral in
Eq.~(\ref{eq41}) is ${\rm {\bf H}}(0,\omega )\,{\rm {\bf n}} = \vert
{\rm {\bf H}}(0,\omega )\vert $ and substituting Eq.~(\ref{eq4})
into Eq.~(\ref{eq41}), we come to the next expression

\begin{equation}
  w_{m\|}  =
   {\frac{{V}}{{80}}}{\frac{{\pi}} {{\Gamma ^{2}}}}\left( {{\frac{{R_{
     \bot}} } {{c}}}} \right)^{2}\vert {\rm {\bf E}}_{0} \vert^{2}{
      \int_{ - \infty} ^{\infty}  {[\sigma (\omega )+\sigma^*(\omega )]\,f(\omega)\,
       \omega ^{2}\;{\frac{{d\omega}} {{2\pi}} }}} .
        \label{eq49}
         \end{equation}

\noindent Using both the Eq.~(\ref{eq42}) and the result of the integration
given by Eq.~(\ref{eq47}), we obtain finally the next expression
for the energy absorbed with this polarization

\begin{equation}
 w_{m\,\|}  =
  {\frac{{1}}{{80}}}{\frac{{V}}{{\sqrt {2\pi} }}}{\frac{{\nu}}
  {{\Gamma}} }\left( {{\frac{{\omega _{p} R_{ \bot} }}{{c}}}}
     \right)^{2}\left( {1 + e^{ - {\frac{{\omega _{0}^{2}}} {{2\Gamma^{2}}}}}}
      \right)\vert {\rm {\bf E}}_{0} \vert ^{2},
       \label{eq50}
        \end{equation}

\noindent provided that $\omega \gg \nu $. The square of
an amplitude of the electrical field in the previous expressions and further
formally could be changed by $\vert {\rm {\bf E}}_{0} \vert ^{2} \to
\,\vert {\rm {\bf H}}_{0} \vert ^{2}$, because the maximal value
of the magnetic field in the pulse would be specified like similar one
in Eq.~(\ref{eq1}), with taking into account that the relation between
fields is given by Eq.~(\ref{eq2}).

If one suppose that a nanoparticle is irradiated by a plane EM wave,
i.e., by the wave (\ref{eq1}) with $\Gamma \to 0$, then the expressions
(\ref{eq48}) and (\ref{eq50}) coincide asymptotically with the known
results from our earlier works\cite{TG,GT}, with an accuracy
up to the constant $2\sqrt {\pi/2}$ [due to the distinction
in a pulse forms (see previous note\cite{foo1})].

\subsection{Kinetic approach}

When the nanoparticle size is less than the electron free pass in
it, the phenomenological approach cannot be used anymore and
ought to be changed by the kinetic approach. Note, that the latter
permits to obtain the correct results for the case when the particle
size is larger than the electron free pass as well. In the kinetic approach,
the current ${\rm {\bf j}}_{ed} (\omega )$ enters into Eq.~(\ref{eq32}),
should be calculated by means of the formula (\ref{eq13}).
Using the nonequilibrium distribution function (\ref{eq33}), one finds

\begin{eqnarray}
 w_{m} &=&{\frac{{4}}{{m}}}{\frac{{e^{2}}}{{R^{4}}}}\left( {{\frac{{m}}{{2\pi
 \,\hbar}} }} \right)^{3}Re{\sum\limits_{i\,j\,k\,l}^{3} {R_{i} R_{j}R_{k} R_{l}}}
  \nonumber \\ &\times&{\int\limits_{ - \infty} ^{\infty}  {{\frac{{
   \alpha_{kj}^{\ast} (\omega )\;\alpha _{il}^{} (\omega )}}{{\nu - i\omega}}}}}
   {\frac{{d\omega }}{{2\pi}} }\int\limits_{ - \infty} ^{\infty} {\upsilon '_{j}
     \upsilon '_{l}\delta (\upsilon ^{2} - \upsilon _{F}^{2} )\,d^{3}\upsilon}
      \nonumber \\ &\times &
      {\int_{V'} {x'_{k} x'_{i} \left( {1 - e^{ - (\nu - i\omega)t'(r',\upsilon ')}}
        \right)\,dr'}} ,
         \label{eq51}
          \end{eqnarray}

\noindent where $\upsilon_{F}$ is the electron velocity on the Fermi
sphere. We have omitted the item in Eq.~(\ref{eq51}) associated with the
second term (with $\upsilon '_{i}$) under the sum sign in Eq.~(\ref{eq33}).
It is not difficult to show that the contribution of this term with
integration over all electron coordinates will be equal to zero, because
it is even with respect to coordinates, whereas an eddy field is an odd
coordinate function. Then, the integration over all coordinates in Eq.~(\ref{eq51}),
associated with the first term under the sum sign in Eq.~(\ref{eq33}),
can be done exactly. We would write here only the final result.
The details of calculations can be found in Ref. \onlinecite{TG}.

 \ba
&&\int_{V'} {x'_{k} x'_{i} \left( {1 - e^{ - (\nu -i\omega )
   \,t'(r',\upsilon ')}} \right)\,dr'}
    \nonumber \\ &=&
     \pi \,R^{5}{\left[{2\psi _{3} {\frac{{\upsilon '_{k}
      \upsilon '_{i}}} {{\upsilon'^{2}}}} + {\frac{{\psi _{1}}} {{2}}}
       \left( {\delta _{ki} - 3{\frac{{\upsilon '_{k}
        \upsilon '_{i}}} {{\upsilon '^{2}}}}} \right)} \right]}.
         \label{eq52}
          \ea

\noindent Here,

\[
 \psi _{1} (\upsilon ',\omega ) = {\frac{{8}}{{15}}} - {\frac{{1}}{{q}}} +
 {\frac{{4}}{{q^{3}}}} - {\frac{{24}}{{q^{5}}}} + e^{ - q}{\frac{{8}}{{q^{3}}}}
   \left( {1 + {\frac{{3}}{{q}}} + {\frac{{3}}{{q^{2}}}}} \right),
    \]

\ba
 \psi _{3} (\upsilon ',\omega ) &=& {\frac{2}{5}} - {\frac{1}{q}}+
 {\frac{8}{3q^{2}}} - {\frac{6}{q^{3}}} + {\frac{32}{q^{5}}}
   \nonumber \\ &-&
    e^{ - q}{\frac{2}{q^{2}}}\left( {1 + {\frac{5}{q}} +
     {\frac{16}{q^{2}}} + {\frac{16}{q^{3}}}} \right),
       \nonumber
        \ea

\begin{equation}
 q = (\nu - i\omega ){\frac{{2R}}{{\upsilon '}}}.
  \label{eq53}
   \end{equation}
The next integration over all velocities space fails to be carried
out in an analytical form for a common case. It can be fulfilled
only if the frequency interval [-$\infty ,\;\infty $]
to split arbitrarily into the two parts

\begin{equation}
 {\int\limits_{ - \infty} ^{\infty}  {\Xi (\omega)\,d\omega =} }2{
   \int\limits_{0}^{\nu _{S}}  {\Xi _{LF} (\omega)\,d\omega}}   + 2{
    \int\limits_{\nu _{S}} ^{\infty}  {\Xi _{HF}(\omega )\,d\omega}}  ,
     \label{eq54}
      \end{equation}
\noindent one of which will exceed, and another one will be less
than the frequency associated with electron oscillations between
particle walls

\begin{equation}
 \nu _{s} = \upsilon _{F} / (2R).
  \label{eq55}
   \end{equation}
\noindent The factor of 2 in Eq.~(\ref{eq54}) appear due to the fact that
the integrants $\Xi(\omega)$ should be the even functions of $\omega $.
The case with $\omega < \nu _{S} $ will be named further as the
lowfrequency (LF) case, and the case with $\omega > \nu _{S} $ --
as the highfrequency (HF) one. Let us consider first the latter.
As a basis for it, we will apply the approximations

\ba
 &&Re{\left\{ {{\frac{{1}}{{\nu - i\omega}} }\psi _{1}
  (\upsilon ',\omega )} \right\}}
    \approx {\frac{{\upsilon'}}{{2R\omega ^{2}}}},
     \nonumber \\ &&
      \quad Re{\left\{ {{\frac{{1}}{{\nu -i\omega}} }
       \psi _{2} (\upsilon ',\omega )} \right\}} \approx
       {\frac{{\upsilon '}}{{8R\omega ^{2}}}},
         \label{eq56}
          \ea
\noindent where

\[
 \psi_{2} (\upsilon ',\omega )=\psi_{3}(\upsilon ',\omega )-
  \frac{3}{4}\psi_{1}(\upsilon ',\omega ).
   \]

\noindent Details of calculations are presented in Refs. \onlinecite{TG} and \onlinecite{GT}.
Here, we write down only the final result which can be obtained
after integration over all electron velocities with the account
of Eqs.~(\ref{eq40}) and (\ref{eq56})

\begin{widetext}
 \begin{eqnarray}
  w_{m,HF} &=& \frac{9}{32}\frac{V}{4\pi}
   \left(\frac{\omega_{p}}{c}\right)^2 \upsilon_{F} R_{\bot}
    \nonumber \\ &\times&
     \int\limits_{\nu_{S}}^{\infty} \left\{
      \eta_{m}^{H} (e_{s} )\,\left( \frac{R_{\|}^{2}}{R_{\vert
       \,\vert}^{2} + R_{ \bot}^{2}} \right)^{2}\vert {\bf H}(0,\omega )
        \vert^{2} + \left[ \rho_{H} (e_{s} ) - \eta_{m}^{H} (e_{s} )\left(
         \frac{R_{\vert \,\vert}^{2}}{R_{\vert \,\vert}^{2} + R_{ \bot}^{2}}
          \right)^{2} \right] \vert {\bf H}(0,\omega )\, {\bf n}\vert^{2}
           \right\}
            \frac{d\omega}{2\pi}\ .
             \label{eq57}
              \end{eqnarray}
               \end{widetext}

\noindent In expression (\ref{eq57}), $\rho _{H} (e_{s} )$
and $\eta _{m}^{H} (e_{s} )$ are the functions depended only on the
spheroid eccentricity $e_{s} $. Their analytical forms are
given in Ref. \onlinecite{TG}. From Eq.~(\ref{eq57}), one can
see that the frequency dependence of the integrand (for any given
polarization) in the HF-case is contained only in the magnetic
field amplitude. Practically this means that we can use the result
of the integration obtained above in the phenomenological approach.
If the lower integration limit in the phenomenological case was the
value of $\nu $, then $\nu _{S} $ will play here the same role.
It is not difficult to ensure that in the case with $\nu _{S} $, one
can diminish the lower limit of integration to the zero as well.
Hence, the energy of a magnetic component of a laser EM wave in
(i) polarization (when the vector of the magnetic component is
directed transverse to the long spheroid axis of revolution),
with an account of the expressions (\ref{eq4}) and (\ref{eq47})
in Eq.~(\ref{eq57}), finally becomes
\ba
 w_{m \bot HF} &=& {\frac{{9}}{{32}}}{\frac{{V}}{{
  \sqrt{2\pi} }}}{\frac{{\omega _{p}^{2}}} {{16\,c^{2}}}}{
   \frac{{\upsilon_{F}}} {{\Gamma }}}\left( {1 + e^{ - {\frac{{
    \omega _{0}^{2}}}{{2\Gamma ^{2}}}}}} \right)
     \nonumber \\ &\times&
      \,\eta _{m}^{H} (e_{s} )\,R_{ \bot}
       \left( {{\frac{{R_{\vert \,\vert }^{2}}} {{R_{\vert
        \,\vert} ^{2} + R_{ \bot} ^{2}}} }} \right)^{2}
         \vert {\rm {\bf E}}_{{\rm {\bf 0}}}\vert ^{2},
          \label{eq58}
           \ea

\noindent and for the (ii) polarization, when, vice verse, the vector
of magnetic field component is directed along this axis, is represented by

\be
  w_{m\,\| HF} =
  {\frac{{9}}{{32}}}{\frac{{V}}{{\sqrt {2\pi} }}}{\frac{{
    \omega_{p}^{2}}}{{16\,c^{2}}}}{\frac{{\upsilon_{F}}}{{\Gamma}}}
     \left({1 + e^{ - {\frac{{\omega _{0}^{2}}} {{2\Gamma ^{2}}}}}}
      \right)\,\rho _{H} (e_{s} )\,R_{ \bot}
       \vert {\rm {\bf E}}_{{\rm{\bf 0}}} \vert ^{2}.
        \label{eq59}
         \ee

Eqs.~(\ref{eq58}) and (\ref{eq59}) coincide each with other for a
spherical particle, because $\eta _{m}^{H} (0) = 4\rho _{H} (0)$.
Besides, if one takes into consideration that the power of the absorbed
energy is $W_m=w_m\Gamma$, where $w_m$ is given by Eqs. (\ref{eq58})
or (\ref{eq59}), and the parameter $\Gamma$ tends to the zero in $W_m$,
then these equations will transform (with an accuracy of the
constant of $2\sqrt{\pi/2}$) into the known ones from our
previous calculations\cite{TG,GT} for a plane wave.

It remains to consider the LF-case. The calculation of Eq.~(\ref{eq51})
in this case can be done with the use of the next approximation

\ba
  &&Re{\left\{ {{\frac{{1}}{{\nu - i\omega}} }\psi _{1}(\upsilon ',
   \omega )} \right\}} \approx {\frac{{R}}{{3\,\upsilon'}}},
    \nonumber \\ &&
     \quad Re{\left\{ {{\frac{{1}}{{\nu - i\omega}} }\psi _{2}
     (\upsilon ',\omega )} \right\}} \approx {\frac{{R}}{{36
       \,\upsilon'}}} .
        \label{eq60}
         \ea

\noindent Furthermore, not complicated but cumbersome
calculations lead us to the next result

\begin{widetext}
 \begin{eqnarray}
   w_{m\,LF} = \frac{3}{16}V\frac{\omega_{p}^{2}} {4\pi
    \,c^{2}}\frac{R_{\bot}^{3}} {\upsilon_{F}}
     \int\limits_{0}^{\nu _{S}}
      \left\{ \eta _{m}^{L} (e_{s} )\left( \frac{R_{\vert \,\vert }^{2}}
       {R_{\vert \,\vert} ^{2} + R_{\bot}^{2}} \right)^{2}
        \vert {\bf H}(0,\omega )\vert^{2} + \left[\rho_{L} (e_{s} ) -
         \eta_{m}^{L} (e_{s} )\left(\frac{R_{\vert
          \,\vert}^{2}} {R_{\vert \,\vert}^{2} + R_{\bot} ^{2}} \right)^{2}
           \right]\vert {\bf H}(0,\omega )\, {\bf n}\vert ^{2}
            \right\}\frac{d\omega} {2\pi} ,
             \label{eq61}
              \nonumber \\
               \end{eqnarray}
                \end{widetext}

\noindent where $\rho _{L} (e_{s} )$ and  $\eta _{m}^{L} (e_{s} )$ are some
smooth functions of the spheroid eccentricity $e_{s} $. The behavior of
these functions depending on the shape of the MN, which is specified by the
ratio of $R_{ \bot} / R_{\vert \,\vert}$, one can find
in Ref \onlinecite{TG}. For a spherical particle

\begin{equation}
 \rho _{L} (0) = \rho _{H} (0) = 2 / 3,
  \qquad \eta_{m}^{L} (0) = \eta _{m}^{H} (0) = 8 / 3.
   \label{eq62}
    \end{equation}
For an estimation of Eq.~(\ref{eq61}), it is necessary to calculate
 the integral

\begin{equation}
 I_{L} = {\int\limits_{0}^{\nu _{s}}  {\omega^{2}f(\omega )\;}} d\omega .
  \label{eq63}
   \end{equation}

\noindent It can be done analytically. It is easy to ensure, however,
that similar to the calculation of the integral $I$ in Eq.~(\ref{eq46}),
one obtains the same result using the integrant from Eq.~(\ref{eq63})
for $I_{L} $. In other words, one can get that

\begin{equation}
  I_{L}\ll I_{\nu _{S}}  .
   \label{eq64}
    \end{equation}
This means that in the sum of Eq. (\ref{eq54}), one can neglect by
the integral of $I_{L} $. In this case, for an arbitrary angle $\theta $
between direction of the magnetic field and the spheroid axis of
revolution, we will obtain finally the expression

\begin{widetext}
 \begin{eqnarray}
   w_{m\,S} = w_{mLF} + w_{mHF}=
    \frac{9}{64}\frac{V}{\sqrt {2\pi}}\frac{\omega_{p}^{2}}{8\,c^{2}}
     \frac{\upsilon_{F}}{\Gamma } \left(1 + e^{-\frac{\omega_{0}^{2}} {2
      \Gamma ^{2}}} \right) R_{ \bot}\left[ \rho_{H} (e_{s} )\sin^{2}\theta +
       \eta_{m}^{H} (e_{s} )\left( \frac{R_{\vert \,\vert}^{2}} {R_{\vert
        \,\vert }^{2} + R_{ \bot} ^{2}} \right)^{2}\cos ^{2}\theta
         \right]\vert {\bf E}_{\bf 0} \vert^{2}.
          \label{eq65}
           \nonumber\\
            \end{eqnarray}
             \end{widetext}
The results (\ref{eq48}) and (\ref{eq50}) obtained in the phenomenological
approach for a spherical nanoparticle go over into the ones
described by Eqs.~(\ref{eq58}) and (\ref{eq59}), or Eq.~(\ref{eq65}),
obtained in the kinetic approach, if one performs formally the next replacement:

\begin{equation}
 {\begin{array}{*{20}c}
  {\nu \to \frac{45}{64} \frac{\upsilon_{F}} {R_{ \bot}}  \left(
  {{\frac{R_{\vert \,\vert} ^{2}} {R_{\vert \,
     \vert} ^{2} + R_{\bot}^{2} }}} \right)\;\eta _{m}^{H} ,
      \qquad \mbox{for}} \, \bot {\mbox{polarization}} \hfill \\
      {\nu \to \frac{45}{32}\frac{\upsilon _{F}} {R_{ \bot}}
        \rho_{H},\qquad \mbox{for}\,
         \vert\vert\mbox{polarization}} \hfill \\
          \label{eq66}
           \end{array}} ,
            \end{equation}

\noindent with the ordinary condition that $\omega\gg\nu $.
In the case of a spherical particle, the above substitution,
obviously, takes the form:

\begin{equation}
  \nu \to {\frac{{15}}{{16}}}{\frac{{\upsilon _{F}}}{{R}}},
   \quad \mbox{provided that}\quad \omega\gg\nu .
    \label{eq67}
     \end{equation}

\subsection{Results and Discussion}

Let us illustrate graphically the expressions obtained
above analytically. We will calculate the ratio
\begin{equation}
  S_{m} = \frac{w_m}{2w}
    \label{eq68}
     \end{equation}

\noindent between the energy absorbed by a unit volume of the MN
and the energy traversed the nanoparticle, given by Eq.~(\ref{eq23}).

Studying the dependence of optical properties of nanoparticles
on their shape, it would be worthwhile to compare the
absorption for particles of various shapes but with equal volumes.
The transformation of a particle shape can be described by the ratio
of radiuses $R_{ \bot} / R_{\vert \,\vert} $. The condition of
fixed particle volume ($V = {\frac{{4\pi}} {{3}}}R_{ \bot}
^{2} R_{\vert \vert}  = const)$ with given ratio of $R_{ \bot} /
R_{\vert \,\vert} $, defines the quantity of $R_{ \bot} $ or of
$R_{\vert \,\vert}$. For instance,
\begin{equation}
 R_{ \bot}  = R\,\left( {{\frac{{R_{ \bot}} } {{R_{\vert
  \vert}} } }} \right)^{1 / 3},\qquad  R_{ \parallel}  = R\,\left( {{\frac{{R_{ \bot}} } {{R_{\vert
  \vert}} } }} \right)^{-2 / 3},
   \label{eq69}
    \end{equation}

\noindent where $R$ is the radius of a sphere of an equivalent volume.

Let us study at first the dependence of $S_{m} $ on the particle
shape. In Figs. $7$ and $8$ for various polarizations of EM wave, the
dependence of $S_{m} $ on the degree of oblateness or prolateness of a
spheroidal particle is depicted (at the frequency of the surface plasmon
$\omega_{0} = \omega _{p} / \sqrt {3} \equiv \Omega $, which was a resonant
one for a spherical particle in the case of the electric absorption).
The calculations for $\bot$ and $\parallel$ polarizations were carried
out using the next formulae
\begin{equation}
 S_{m\bot}=\frac{\nu}{10\pi}\Omega^2 \left(\frac{R}{c}\right)^3
  \frac{(R_{\bot}/R_{\parallel})^{2/3}}{1+(R_{\bot}/R_{\parallel})^2},
   \label{eq70}
    \end{equation}
\begin{equation}
 S_{m\parallel}=\frac{\nu\Omega^2}{20\pi}\left(
  \frac{R}{c}\right)^3 (R_{\bot}/R_{\parallel})^{2/3},
   \label{eq71}
    \end{equation}
for phenomenological case, and

\begin{equation}
 S_{m\bot}=\frac{9}{128\pi}v_F\Omega^2 \frac{R^2}{c^3}\eta^H_m (e_s)
  \frac{(R_{\bot}/R_{\parallel})^{1/3}}{[1+(R_{\bot}/R_{\parallel})^2]^2},
   \label{eq72}
    \end{equation}

\begin{equation}
 S_{m\parallel}=\frac{9}{128\pi}v_F\Omega^2
  \frac{R^2}{c^3}\rho_H (e_s)(R_{\bot}/R_{\parallel})^{1/3},
   \label{eq73}
    \end{equation}
for kinetic one.

The calculation are performed assuming the next numerical parameters
for Au particle: the electron concentration\cite{CK} of
5.9$\times10^{22}$ cm$^{ - 3}$, $\Omega = 7.91\times 10^{15}$ s$^{-1}$,
$\nu = 3.39\times 10^{13}$ s$^{ - 1}$,
and $\upsilon_{F} = 1.39 \times 10^{8}$ sm/s.

\vskip10pt
\noindent\includegraphics[width=8.6cm]{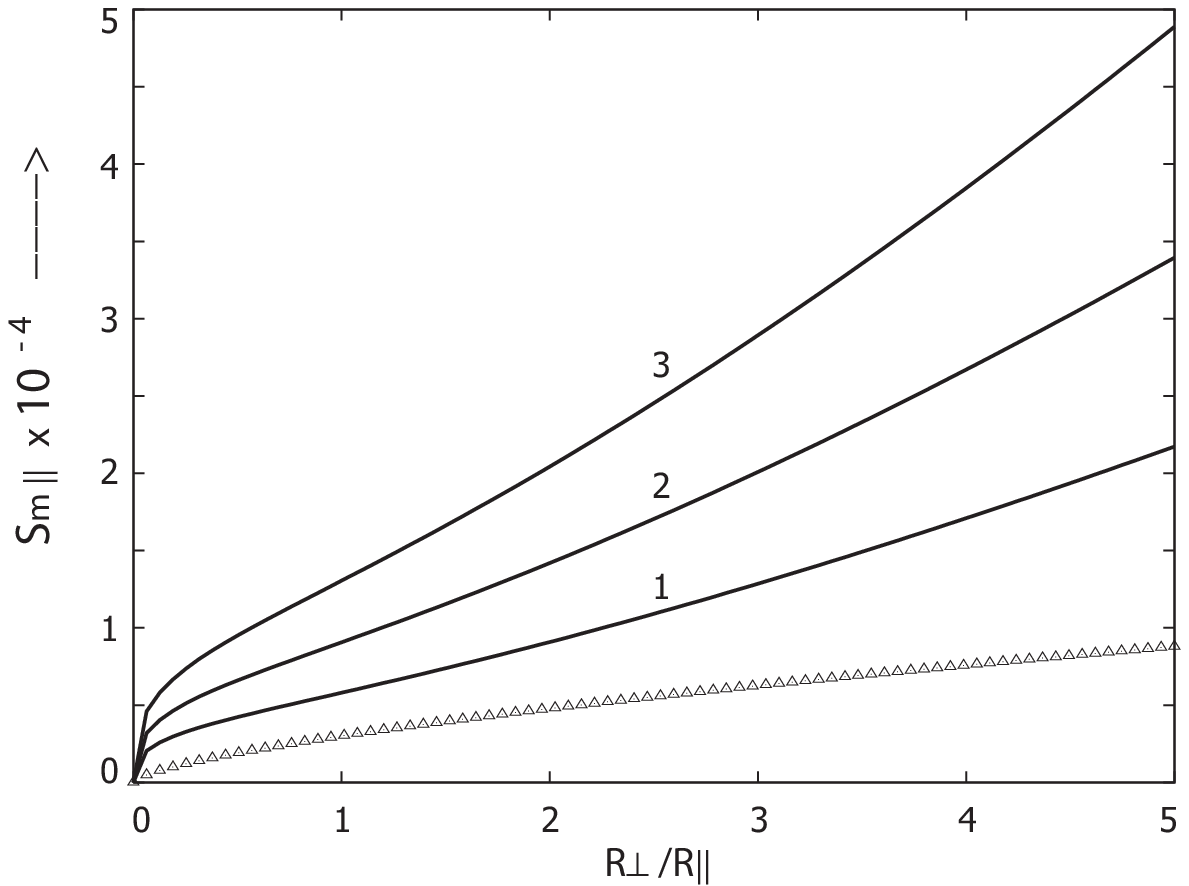}

\vskip-1mm\noindent{\footnotesize FIG. 7. The energy
absorbed by Au particle for polarization of a magnetic field
transverse ($ \parallel )$ to it axis of revolution vs degree of spheroid
oblateness or prolateness. Curves are obtained at the frequency of
the plasmon resonance $\omega _{0} =\Omega $, for different
$R$, \AA: 200 (curve $1)$, 250 (2), 300 (3). The phenomenological
dependence for $R$=200 \AA is given by triangles.}%
\vskip10pt

The curves $1-3$ correspond to the different particle radiuses, whereas
the curve of phenomenological dependencies builded on the base of the
formulae (\ref{eq70}) and (\ref{eq71}) is given by triangles for $R =200$ \AA.
Comparing curves $1-3$, it is clearly seen that the energy of laser
pulses will be absorbed more strongly in Au particles of a larger volume.
When the magnetic field polarization is directed along the MN rotation axis
($\vert \vert$ polarization), the absorption grows with enhancing of the
particle oblateness (Fig. 7), while for the transverse polarization
of the magnetic field ($ \bot $ polarization), the absorption reaches the
maximum for some values of the ratio of $R_{ \bot}  / R_{\vert\,\vert}$ (Fig. 8).
In the phenomenological case, this maximum is reached for particles
of a prolate shape when $R_{\bot}/R_{\parallel}=1/\sqrt {2}$, and
in the kinetic one -- when $R_{ \bot} / R_{\vert \,\vert}
= 1 /\sqrt {11}$. If one compare the magnitude of the absorption
maximum for the same pulse duration at $\bot $ polarization, which
has been obtained in the phenomenological and the kinetic approaches
from formulae (\ref{eq48}) and (\ref{eq58}),
then one can get the relation
\begin{equation}
 {\left. {{\frac{{w_{m \bot ,HF}}} {{w_{m \bot}} } }}
   \right|}_{\max} \simeq {\frac{{4}}{{5}}}\left( {{\frac{{
    \upsilon_{F}}}{{\nu \,R}}}} \right).
     \label{eq74}
      \end{equation}
\vskip10pt
\noindent\includegraphics[width=8.6cm]{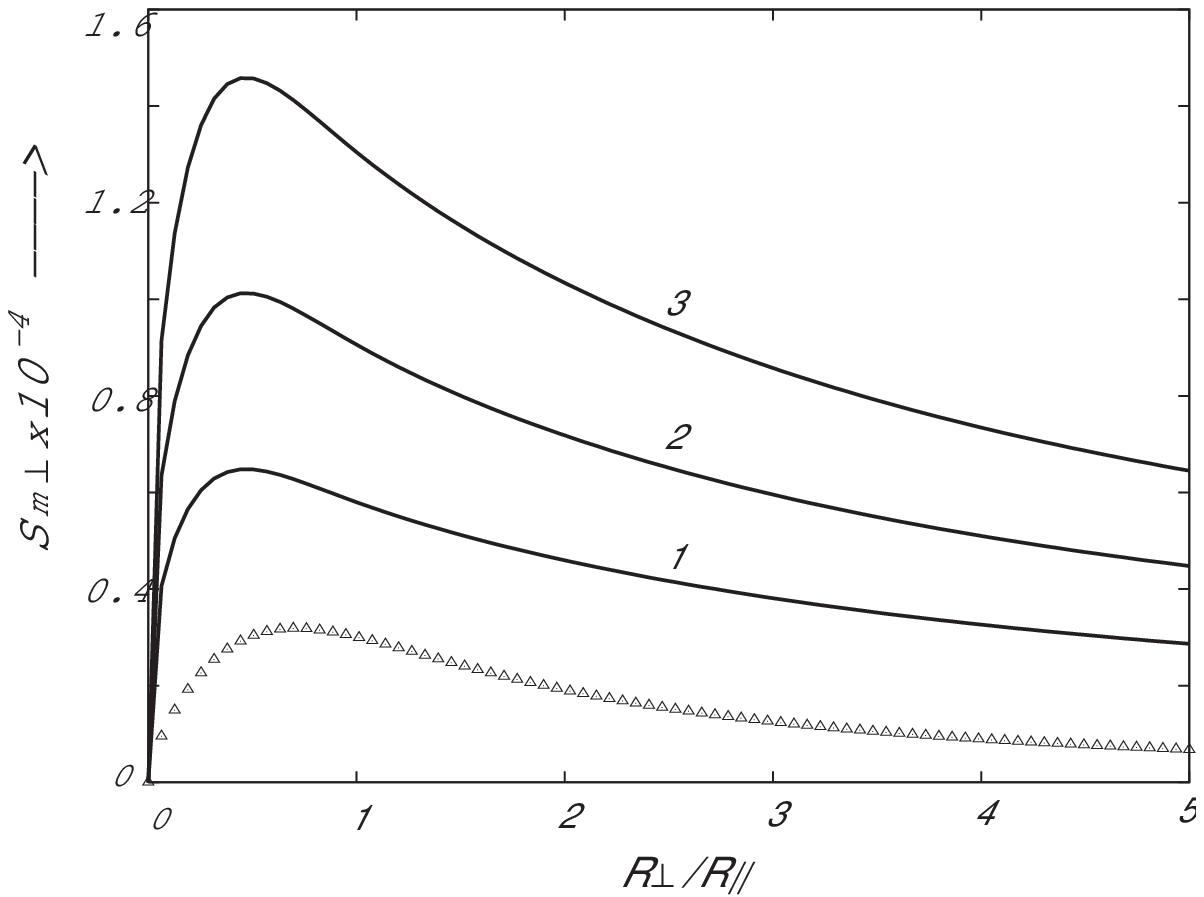}

\vskip-1mm\noindent{\footnotesize FIG. 8. The same, as in Fig. 7
for \textit{$\bot$} polarization of magnetic field.}
\vskip10pt

From Eq.~(\ref{eq74}) it follows that this relation increases with the
reduction of a particle radius and does not depend on laser beam parameters.
The same one can see from Eqs. (\ref{eq70})--(\ref{eq73}), where the ratio
between absorbed and traversed energy in the case of magnetic absorption
does not depend on parameter $\Gamma$. For Au particle with the radius, for
instance, 200 {\AA} one can see (comparing curves 1 and those labeled by
triangles in Fig. 8) that the absorption in the kinetic case is
approximately of two times of magnitude higher.

With the extension of a MN prolateness, the magnitude of the
absorption is decreased in the case of $ \bot $ polarization of the
magnetic field, and increased -- in the case of $\vert \vert$~polarization.
The reason for this is that the number of electron closed orbits
is increased with growing of a prolateness of a particle in
$\vert \vert$~polarization, and decreased in
$ \bot $~polarization of the magnetic field.

Contrary to the electric absorption, in the dependence of the absorbed
energy on $R_{\bot} / R_{\vert \,\vert}$, there does not occur neither
the shift of maximum (for $ \bot $~polarization) nor an appearance of
any others maxima with the deviation of a carrier frequency
from the frequency of the surface plasmon $\Omega$.

For the case when $\omega_{0} < \Omega$, the intensity of magnetic
absorption slightly increases with decrease of $\omega_0$ and
asymptotically approaches the constant value, when
$\omega_{0} > \Omega$ [see Eqs. (\ref{eq58}),
(\ref{eq59}), (\ref{eq48}), and (\ref{eq50})].
This is true for both polarizations as well as for both
the phenomenological and the kinetic approaches. But depending
on the duration of an incident pulse, a certain peculiarities
appears. Namely, the absorption intensity is increased with
an enhancement of the pulse duration at different carrier frequencies
for both examined here polarizations. This grows is more
rapid at frequencies $\omega_{0}\ll\Omega $, but has a restriction
connected with the evident fact that the space size of a pulse
$c / \Gamma $ cannot exceed the length of the carrier wave $\lambda _{0} $
in vacuum: $\Gamma / \omega _{0} \ge 1 / (2\pi ) \approx 0.16$.

In contrast to spherical particles, the magnetic absorption in
nonspherical ones drastically depends on magnetic field polarization.
So, the absorption of a spheroidal MN increases for moderate prolate
and decreases for oblate particles (comparing to the particles of a
spherical shape of the same volume) at $ \bot $~field polarization.
It is the least for particles of an oblate shape. In the case of
$\vert \vert $~polarization, vice versa: the absorption noticeably
increases for an oblate and decreases for a prolate MN comparing to
the spherical one. Similar trends holds true for phenomenological
approach in both polarization as well.

If one choose a nanoparticle of a spheroidal shape with some fixed
ratio of $R_{ \bot} / R_{\vert \,\vert} $ and will change only the
magnitude of the carrier frequency, then one can establish using
Eqs. (\ref{eq58}), (\ref{eq59}), (\ref{eq48}), and (\ref{eq50})
for different polarizations of an incident field the following.
(i) For MN of a prolate shape ($R_{\bot}/R_{\vert\,\vert}<1$) with
an enhancement of a laser pulse duration at frequencies
$\omega_{0}\ll\Omega $, the maximum of the energy absorption increases
and does not split into the two peaks as we obtain above for prolate Au
particle in an electric fields with $\Gamma $ exceeding some limit
quantity (see Figs. 5 and 6). (ii) If the ratio of $\omega_0 /\Omega$ is small,
the rise of magnetic absorption reaches the higher point for the least
frequency $\omega_0 = \nu _{S} $. (iii) Comparing intensity of absorption
for pulses of equal duration, but different polarization, one can
ensure that more rapid enhancement of the energy absorption by
MN with increasing of carrier frequency occurs at the $ \bot $~polarization
of the magnetic field.

For oblate particles, vice verse, an enhancement of the energy absorption
by MN with decreasing of a carrier wave proceeds faster for the $\vert \vert
$~-polarization of the magnetic field, then for it $ \bot $~polarization.
The pulses of larger duration (with small value of $\Gamma )$ are intensively
absorbed at lower frequencies of a carrier wave in both
polarizations.

Thus, the spheroidal MN of an oblate or prolate shape comparing to
the spherical one can absorb the energy of the magnetic field
from ultrashort laser pulses more or less intensively depending
on the magnetic field orientation with respect to the spheroid axis
of revolution.

It would be interesting also to monitor the absorption by a
nanoparticle of a fixed shape with volume changing. This can be
easy fulfilled for fixed ratio of $R_{ \bot} / R_{\vert\,\vert} $ with
changing the radius of an equivalent sphere $R$ in the previously obtained
formulae. Such one dependence for nonspherical particles can be strongly
differ from an analogous one in the case of spherical particles and
requires a special investigation. There is only problem, when the volume
of particle is changed with simultaneous preservation of its shape, the
size of a particle in same directions becomes greater than the electron free
pass in those directions. Similar study for an electric absorption we have
performed in Ref. \onlinecite{GT4}.

From expressions (\ref{eq42}) and (\ref{eq52}), or (\ref{eq44})
and (\ref{eq53}) one can see only that the energy of the magnetic
absorption with the use of the phenomenological description increases
quadratically with $R_{ \bot} $, whereas at the kinetic description,
the growth proceeds linearly with the $R_{ \bot} $. Generally, the
increase of magnetic absorption with $R_{ \bot}$ is due to the
enhancement of the number of electron closed orbits with growing of
a particle oblateness in both polarization of the magnetic field.
The different behavior of the two approaches with $R_{ \bot}$ can be
understood from above Eq. (\ref{eq66}), where phenomenological
parameter $\nu$ is replaced by $v_F/R_{\bot}$ in a kinetic approach.

By making use of the above results, namely, Eq. (\ref{eq19}) for the
electric absorption and Eqs. (\ref{eq58}) and (\ref{eq59}) valid for the
magnetic absorption in a high-frequency case, one can compare
relative contributions for both polarizations.

\ba
 \left(\frac{w_m}{w_e}\right)_{||,\bot} &\simeq& {\frac{9}{16
  \sqrt{2\pi}}}{\frac{\upsilon_{F}}{c}}
   \left(\frac{{\Gamma}R_{\bot}}{c}\right)(1+e^{-
    \frac{\omega^2_0}{2\Gamma^2}}){e^{ {\frac{{
    (\omega_{||,\bot}-\omega _{0})^{2}}}{{2\Gamma ^{2}}}}}}
     \nonumber \\ &\times&
      \left\{
       \begin{array}{*{20}c}
        \rho_H (e_{s}) \hfill  \\
         \eta _{m}^{H} (e_{s} )
          \left( {{\frac{{R_{\vert \,\vert }^{2}}} {{R_{\vert
           \,\vert} ^{2} + R_{ \bot} ^{2}}} }} \right)^{2} \hfill.
            \end{array}
             \right.
              \label{eq75}
               \ea
For spherical particle, using Eq. (\ref{eq62}), one can get that

\begin{equation}
 \left(\frac{w_m}{w_e}\right)_{sph} \simeq {\frac{3}{8
  \sqrt{2\pi}}}{\frac{\upsilon_{F}}{c}}
   \left(\frac{{\Gamma} R}{c}\right)(1+e^{-\frac{\omega^2_0}{2\Gamma^2}})
    {e^{ {\frac{{(\omega_{p}/\sqrt{3} - \omega _{0})^{2}}}{{2\Gamma ^{2}}}}}}.
      \label{eq76}
       \end{equation}
An estimation of the ratio of absorptions [from Eq. (\ref{eq76})] for
particle with the radius of $R=50 \AA$, for which the input of the
magnetic field overcomes the electric one for a separate
absorption,\cite{GT4} shows that at the frequency $\omega_0=\Omega$
the absorption associated with an electric plasmon excitation is considerably
higher than the magnetic one. The same one can see, comparing the magnitudes
of absorption shown in Figs. 7 and 8 for magnetic absorption, with electric one
(Fig. 2) for Au particle at plasmon resonances. However, the ratio between
magnetic and electric fields input can enhanced with the pulse duration, and
for plane monochromatic wave will be equal to 1 or even exceed 1, depending on the
particle sizes. The substantial rice of this ratio at plasmon resonances can
be realized only in $\parallel$ polarization of magnetic field for highly oblate
particles. For instance, for Au particle of $R=200$\AA, we can reach
the same input of magnetic component as for electric one (with
$\Gamma =2.637\times 10^{15}$s$^{-1}$) at $R_{\bot}/R_{\parallel}\approx 800$.
The input of a magnetic component in the absorption becomes comparable and even
can exceed the electrical one, as well, if a carrier frequency
$\omega_0$ deviates from the plasmon frequency $\Omega$, especially at small
values of $\Gamma$. One can find from Eq. (\ref{eq76}) that, e.g., for Au
particle with $R=200$\AA, and $\Gamma=1.13\times 10^{15}$ s$^{-1}$, at
$|\Omega -\omega_0|=5\times 10^{15}$ s$^{-1}$,  ${w_m}/{w_e}\simeq 1$.

In general, the absorption of energy from ultrashort laser pulses by spheroidal MN
depends both on the particle and the pulse characteristics. If the parameters
of laser pulses are given, the absorption by MN depends on its volume (for fixed shape)
as well as on the its shape (for fixed volume). We have especially interested
in the dependence on the particle shape, because with changing of it, we can variate
the position of plasmon resonances on the frequency scale. In other words, with
changing the ratio of $R_{\bot}/R_{\parallel}$, the plasmon resonances may came
close to or move away from the fixed carrier frequency.
That is important for the amount of energy absorbed by particle. Such an approach
is very demonstrable and productive for a theoretical studding of the optical
properties of a MN. But for direct comparison of a theoretical predictions with an
experiments it is necessary to have the data for a metallic nanoparticles of equal
volumes but different in their shapes.

\section{CONCLUSIONS}

We develop the theory of the absorption of ultrashort laser pulses
of a different duration and carrier frequencies by small metallic
particles of a spheroidal shape. The cases when an electron free
pass is greater or much less than a particle size are considered.
There has been analyzed the dependence of the absorbed energy on the
degree of a particle oblateness or prolateness at the frequency of a
plasmon resonance and at frequencies that are greater or less than it.
The simple analytic expressions are obtained, which make it possible
to study the plasmon and magnetic absorption of laser pulses depending
on the duration of these pulses, the magnitude of the carrier frequency,
the particle shape, and the field polarizations. The analysis of the dependence
of the energy absorbed by a spheroidal MN on the degree of its prolateness
or oblateness was carried out for different values of pulse duration and
frequencies that are higher, lower, or equal to the plasmon
resonance frequency.

At the frequency of a carrier wave, which is equal with that of a
surface plasmon, the maximum absorption was observed for spherical
MNs. As soon as the carrier frequency deviates from that of the
surface plasmon in a spherical particle, the two maxima appear in the
dependence of the absorbed energy on the ratio of spheroid semiaxes:
one of them corresponds to the prolate particles, while the other --
to oblate particles. As the frequency deviates from the
resonance one, the peak of the absorbed energy at first decreases
in an absolute value, then splits and finally stabilizes for particles
being more and more prolate or oblate.

For the case of spheroidal MNs subjected to the laser irradiation,
with increase in the pulse duration, the peak in the dependence of
the absorption on the frequency of a carrier wave also splits into
two peaks located on opposite sides of the resonance (for a spherical
particle) frequency. The double
peaks originate from the resonances excited by ultrashort laser
pulses at the frequencies of plasmon oscillations along or transverse
to the spheroid rotation axis. A distance from the minimum in a well
between the doublet peaks to the frequency of a plasmon resonance
for a spherical particle provides the information about the degree
of prolateness or oblateness of a nanoparticle.

As the degree of prolateness or oblateness increases, the distance
between the doublet components grows. At the fixed incident angle of
a laser pulse, the peak height keeps constant, with its value being
dependent only on the pulse duration. It was considered how the change
in the incident angle of a laser pulse directed along or transverse
to the spheroid rotation axis effects the value of the relative
absorption intensity.

We have found that the absorption in nonspherical particles drastically
depends on the magnetic field polarization with respect to the particle
rotation axis. Namely, for MN of a spheroidal shape, with the direction
of the magnetic field transverse to the spheroid axis of revolution
($ \bot $~polarization), the absorption is increased for a prolate and
decreased for an oblate MN compared to a particle of a spherical shape
of the same volume. The case of the magnetic field orientation along the
axis of revolution of a particle ($\vert \vert $~polarization) proved
to be the contrary one: the absorption is increased for a oblate MN and
decreased for a prolate one.

It appears that the energy of laser pulses with the larger value
of $1/\Gamma$ is absorbed by MN better than with the smaller one.
In the case with increasing of a particle \textit{oblateness}, the magnetic
absorption is increased for longitudinal ($\vert \vert )$, and decreased --
for a transverse ($ \bot )$ polarization of the magnetic field.
Comparing to the MN of spherical shape, the magnetic absorption
for nanoparticles of a \textit{prolate} shape is increased with
rising of particle prolateness at $ \bot$~polarization of magnetic
field, reaching maximum for the ratio of spheroid halfaxes
$R_{ \bot}  / R_{\vert \,\vert}\approx 7 / 15$, and diminishes
for the larger prolateness. At $\vert \vert $~polarization,
increasing in a prolateness leads to decrease in the absorption.

In the kinetic approach, the maximum value of the magnetic absorption
for a fixed pulse duration and at $ \bot$~polarization of magnetic
field can exceed by two times or more (for particles with $R \leq 200 ${\AA})
the absorption, obtained in the phenomenological approach [at $T=0^o$C].

With reduction in a carrier frequency, the intensity of the magnetic absorption
by MN of a prolate shape is increased for $ \bot$~polarization of the magnetic
field more distinctly than at it $ \parallel $~polarization.
As soon as it took place, the pulses with longer duration are absorbed
better at the lowest frequencies of the carrier wave. For oblate particles,
on the contrary, the faster enhancement of the energy absorption by MN with
increasing carrier frequency occurs for $\vert \vert$~polarization
of the magnetic field, than at it $ \bot $~polarization.

For frequencies of carrier wave close to the plasmon resonances in spheroidal
MN, the electric absorption is considerably higher than the magnetic one at
$\bot$ polarization of magnetic field and can be comparable with one --
at $\parallel$ polarization for highly oblate MN.

A comparison between the results obtained in both the
phenomenological and kinetic approaches have been advanced.


\begin{thebibliography}{50}
\bibitem[24]{foo1}   This dependence appears in Eq.~(\ref{eq16}) from $\gamma$
                     which depends on $R$ due to the conductivity tensor (\ref{eq18}).
\bibitem[26]{foo2}   The multiplier $1/2$ appears as a result of different
                     choices of the expressions for the EM wave field: in work \onlinecite{TG},
                    we set it as $\mathrm{\mathbf{E}}_{0}\exp [i(\omega
                    t-\mathrm{\mathbf{k}}\mathrm{\mathbf{r}})]$;
                    but, here at $\Gamma \rightarrow 0$, we have $\mathrm{\mathbf{E}}%
                    _{0}\cos (\omega _{0}t-\mathrm{\mathbf{k}}_{0}\mathrm{\mathbf{r}})$
                    according to formula (\ref{eq1}). The another multiplier
                    $\sqrt{\pi /2}$ results from the fact that we deal here with the
                    Gaussian-like, rather than rectangular pulse.
\end{thebibliography}
\end{document}